\documentclass[%
reprint,
superscriptaddress,
 amsmath,amssymb,
 aps,
prb,
]{revtex4-2}
\usepackage{xcolor}
\usepackage{feynmp-auto}
\usepackage{graphicx}
\usepackage{dcolumn}
\usepackage{bm}
\usepackage{amsmath}
\usepackage{amssymb}
\usepackage[svgnames]{xcolor}
\usepackage{mathtools}
\usepackage{braket}
\usepackage{physics}
\usepackage{subfiles}
\usepackage{booktabs}
\usepackage{multirow}
\usepackage[colorlinks=true,citecolor=blue,linkcolor=blue]{hyperref}
\usepackage[compress]{cleveref}

\newcommand{\eqgraph}[3]{%
  \begin{gathered}
  \raisebox{0pt}[\dimexpr\height+#1][\dimexpr\depth+#2]{\ignorespaces#3\unskip}%
  \end{gathered}
}

\newcommand{\IR}{\text{IR}}
\newcommand{\R}{\text{R}}
\newcommand{\PhIR}{\phi_\text{IR}}
\newcommand{\PhR}{\phi_\text{R}}
\newcommand{\ret}{\text{ret}}

\newcommand{\eg}{\text{E$_{2g}$} }
\newcommand{\eu}{\text{E$_{1u}$} }
\newcommand{\inv}{^{\raisebox{.2ex}{$\scriptscriptstyle-1$}}}
\newcommand{\out}{\text{out}}
\newcommand{\q}{\textbf{q}}
\newcommand{\qp}{{\textbf{q}'}}
\renewcommand{\k}{\textbf{k}}

\newcommand{\emb}{\mathrm{em}}
\newcommand{\fp}{\mathrm{f.p.}}

\renewcommand{\adv}{\text{adv}}
\def\code#1{\texttt{#1}}

\begin{document}

\preprint{APS/123-QED}

\title{Probing quantum nonlinear phononics in multilayer graphene with time-resolved Raman scattering
}

\author{Jakob Dolgner}
\email{j.dolgner@fkf.mpg.de}
\affiliation{%
 Quantum Many-Body Theory Department, Max Planck Institute for Solid State Research, Stuttgart, Germany
}
\author{Michael Rübhausen}
\email{michael.ruebhausen@uni-hamburg.de}
\affiliation{
 Institute of Nanostructure
and Solid State Physics, Hamburg University, Hamburg, Germany
}%
\author{Dirk Manske}%
\email{d.manske@fkf.mpg.de}
\affiliation{%
 Quantum Many-Body Theory Department, Max Planck Institute for Solid State Research, Stuttgart, Germany
}%

\begin{abstract}
Coherently driven phonons are conventionally described classically. Nonlinear phonon–phonon interactions, however, allow an infrared drive to reshape the fluctuation spectrum and generate substantial nonthermal enhancement of phonon fluctuations, motivating a description beyond the coherent displacement alone. We benchmark classical and semiclassical approximations against the full quantum theory and find that the semiclassical treatment substantially improves agreement with the quantum dynamics. We propose nonequilibrium spontaneous Raman scattering as a sensitive probe of these effects and apply our framework to recent experiments suggesting quantum nonlinear phononics involving the \eu and \eg modes in multilayer graphene. At very strong pump fields, driving the \eu mode significantly modifies the \eg Raman spectrum, however reproducing the reported response at experimental field strengths requires accounting for pump-induced Raman matrix element dynamics or alternative energy pathways into the \eg mode. Our findings extend nonlinear phononics beyond average atomic displacements to the dynamics of lattice fluctuations, highlighting the interplay between coherent motion and quantum correlations in driven materials.
\end{abstract}


\maketitle

\date{\today}

\section{Introduction}

Selective pumping of phonons offers a controlled means of driving a solid away from equilibrium. A resonant mid-infrared (MIR) or terahertz (THz) frequency field can generate a macroscopic amplitude of an infrared (IR)-active mode. Nonlinear phononics describes how anharmonic couplings convert this motion into rectified or oscillatory forces on other lattice modes \cite{merlin_generating_1997,forst_nonlinear_2011,subedi_theory_2014,forst_mode-selective_2015,neugebauer_comparison_2021}. This principle has been explored to induce a broad range of collective phases. Mode-selective vibrational excitation can melt orbital order in manganites \cite{rini_control_2007,tobey_ultrafast_2008}; reverse, induce, or spatially transport ferroelectric polarization \cite{mankowsky_ultrafast_2017,nova_metastable_2019,li_terahertz_2019,henstridge_nonlocal_2022}; and generate effective magnetic fields, modify exchange interactions, polarize antiferromagnets, switch magnetization, or stabilize high-temperature ferromagnetism \cite{nova_effective_2017,disa_polarizing_2020,afanasiev_ultrafast_2021,stupakiewicz_ultrafast_2021,disa_photoinduced_2023,basini_terahertz_2024}. 

In the cuprates, resonant lattice excitation has been associated with transient interlayer coherence, amplified superconducting fluctuations, and magnetic-field expulsion \cite{fausti_light_2011,hu_optically_2014,mankowsky_nonlinear_2014,liu_pump_2020,vonhoegen_amplification_2022,fava_magnetic_2024}. In K$_3$C$_{60}$, superconducting-like optical and nonlinear-transport signatures exhibit metastability, pressure dependence, and strong resonances with intramolecular vibrations \cite{mitrano_possible_2016,cantaluppi_pressure_2018,budden_evidence_2021,rowe_resonant_2023,wang_transport_2023}. These observations make phonon-driven phase control a compelling route toward states inaccessible in equilibrium, although the microscopic balance between coherent lattice dynamics and other driven electronic processes remains material dependent. 
Unlike conventional electronic photoexcitation, this approach acts primarily through a chosen structural coordinate and can therefore modify the electronic energy landscape without first producing a dense population of incoherent carriers \cite{rini_control_2007,tobey_ultrafast_2008}. Consequently, the lattice is promoted from a passive heat bath to an active control parameter for nonequilibrium order.

Conventional theoretical treatments of phonons in these phenomena are classical. The relevant normal coordinates are represented by classical fields obeying coupled, damped equations of motion in an anharmonic potential, whose parameters are commonly extracted from frozen-phonon density-functional calculations \cite{merlin_generating_1997,subedi_theory_2014,fechner_effects_2016,juraschek_ultrafast_2017,radaelli_breaking_2018,mankowsky_nonequilibrium_2016,disa_engineering_2021}. This approach is natural for the coherent amplitude: a spatially uniform field drives the zone-center mode macroscopically, so that its relative quantum uncertainty vanishes in the thermodynamic limit. It does not follow, however, that observables of the driven state evolve according to their classical equations of motion. 
Anharmonic interactions couple the coherent displacement to phonon fluctuations. On the one hand, this allows parametric generation of squeezed states \cite{garrett_ultrafast_1997}; on the other hand, back-action from the fluctuation dynamics influences the evolution of the coherent amplitude. Phonon fluctuations have been investigated using femtosecond X-ray diffraction, optical photon statistics, and time-resolved diffuse scattering \cite{johnson_directly_2009,esposito_photon_2015,fechner_quenched_2024}; quantum treatments of impulsive Raman and pump-probe spectroscopy show that they contain information absent from the mean trajectory \cite{glerean_quantum_2019,lakehal_detection_2020,libbi_quantum_2025}. Quenching quantum fluctuations via nonlinear phononics enables a laser-driven transition from paraelectricity to ferroelectricity in SrTiO$_3$~\cite{fechner_quenched_2024,libbi_laser-driven_2026}.

In Sec.~\ref{sec:quantum_nonlinear_phononics}, we show that fluctuation corrections to driven nonlinear phononics remain independent of the thermodynamic limit. For the paradigmatic cubic interaction $Q_\R Q_\IR^2$, which we use as an example, the equations of motion for both the mean phonon amplitudes and their two-point correlation functions contain fluctuation contributions that scale extensively, just like the coherent forcing. Moreover, these contributions acquire genuine pump-induced dynamics and therefore cannot, in general, be absorbed into equilibrium-renormalized phonon parameters. This establishes the connected phonon correlator, capturing the fluctuation spectrum, as an essential observable of the driven state.
\begin{figure}
    \centering
    \includegraphics[width=0.99\linewidth]{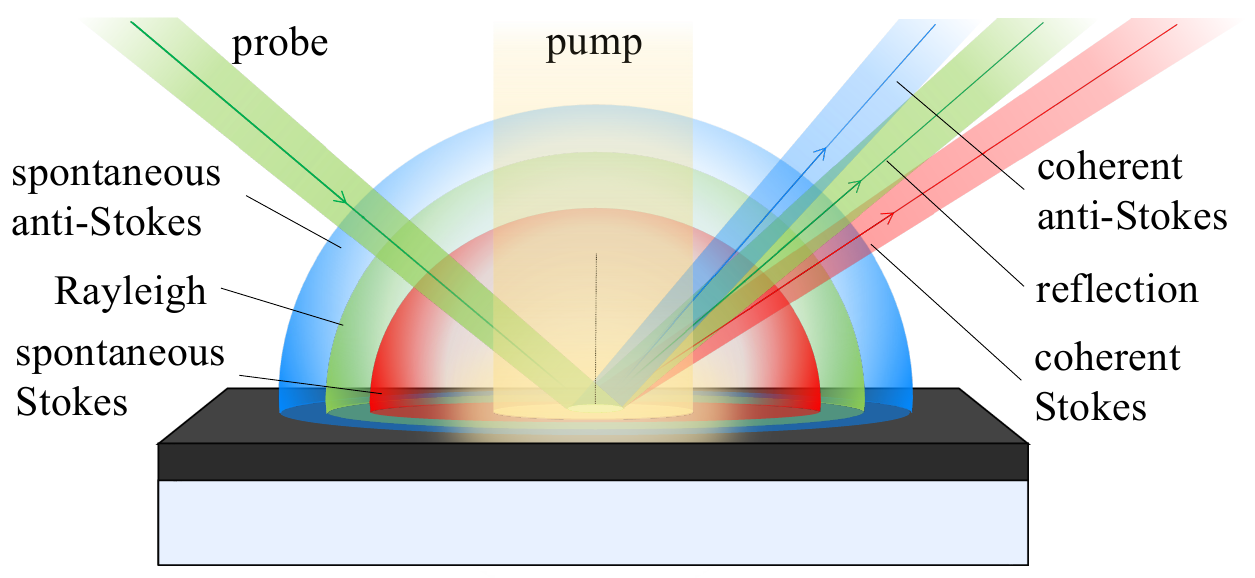}
    \caption{Illustration of the different scattering channels found in nonequilibrium phononic Raman scattering. The backward-scattering direction of the coherent Stokes and anti-Stokes processes is not shown for clarity.}
    \label{fig:Raman_sketch}
\end{figure}

In Sec.~\ref{sec:Nonequilibrium_Raman_theory}, we show that time-resolved spontaneous Raman scattering provides direct access to this observable. The nonequilibrium Raman intensity is governed by the full greater phonon correlator, which separates into a connected contribution from quantum and thermal fluctuations and a disconnected contribution from the coherent phonon amplitude. These two terms are not only conceptually distinct; their radiation patterns differ, as Fig.~\ref{fig:Raman_sketch} illustrates. The coherent contribution is emitted into sharply selected directions set by phase matching, whereas the fluctuation contribution observed in spontaneous Raman scattering remains broadly distributed. Their angular separation therefore offers a practical way to isolate the driven fluctuation spectrum and distinguish it from coherent Raman sidebands. This complements established time-resolved Raman approaches to nonequilibrium electronic excitations \cite{wang_theory_2018,werner_nonequilibrium_2023,shvaika_interpreting_2018}.

In Sec.~\ref{sec:Semiclassical} we derive a semiclassical approximation that accounts for fluctuations but neglects incoherent scattering. We compare it with the classical approach and the quantum dynamics of the specified local phonon model under a Lindblad master equation in the short-but-intense-pumping regime. Compared to the classical approach, the semiclassical approximation agrees much better with the full quantum evolution of the coherent phonon amplitude while having identical numerical complexity. Furthermore, the semiclassical dynamics of the fluctuation spectrum agree well with the exact one. 

As a concrete application, in Sec.~\ref{sec:graphite} we analyze the nearly degenerate IR-active \eu and Raman-active \eg optical phonons of AB-stacked multilayer graphene. A recent pump--probe experiment found that resonantly driving the \eu mode shifts the \eg Raman peak toward the pump frequency: the mode softens for red detuning and hardens for blue detuning analogously to the optical spring effect of cavity optomechanics \cite{buchenau_optically_2023,aspelmeyer_cavity_2014}. This behavior was interpreted as a nonequilibrium level-repulsion effect, but its microscopic origin remained unresolved. Combining a quantum master-equation treatment of the phonons with ab initio estimates of anharmonic couplings, we find that phonon anharmonicities dynamically modify the fluctuation spectrum. However, a semiclassical/Gaussian approximation suffices to capture these modifications qualitatively. Furthermore, phonon interactions alone cannot account for the observed shift. Instead, pump-induced modification of the Raman tensor may play a central role. Previous observations of drastically enhanced electron-phonon coupling upon driving the \eu mode in bilayer graphene \cite{pomarico_enhanced_2017} support this hypothesis. Our results thus identify spontaneous Raman scattering as a sensitive probe of quantum nonlinear phononics but caution that pump-induced spectral changes need not arise solely from intrinsic phonon anharmonicity.

\section{Quantum nonlinear phononics}\label{sec:quantum_nonlinear_phononics}
Let us briefly review the classical phononics approximation.
Consider a quantum field of mass-weighted normal coordinates $Q(\mathbf{r},t)$ describing the normal mode of a harmonic optical phonon, where $\mathbf{r}$ denotes the unit-cell coordinate in configuration space. Its Lagrangian is that of simple harmonic oscillators
\begin{align}\label{eq:harmonic_phonon}
    L(t) &=  \frac{1}{2}\sum_\q \varphi_{-\q}(t) G\inv_\q \varphi_\q(t) 
\end{align}
with the inverse propagator $G\inv_\q = - (\partial_t^2 + \omega_\q^2)$ (we use $\hbar = 1$) and
$\varphi(\q,t) = 1/\sqrt{N}\sum_\textbf{r} Q(\textbf{r},t) e^{-i\q\cdot\textbf{r}}$,
where $N$ denotes the number of unit cells. The fields have dimension $[\varphi] = \textit{energy}^{-1/2}$.
Assuming the phonon is IR-active, a macroscopic external field can interact linearly with the phonons according to
\begin{equation}\label{eq:external_driving_realspace}
    V_\mathrm{l-ph} =  -\sum_\textbf{r} Q(\textbf{r},t) F(t)\,,
\end{equation}
with the force
\begin{equation}
    F(t) = e\int_{\omega} e^{-i\omega t} \mathbf{d}(\omega)\cdot \mathbf{E}(\omega)
\end{equation}
where we have used the dipole approximation for the optical electric field \textbf{E} and $\textbf{d}$ denotes the oscillator strength, that is the mass-weighted Born effective charge of the phonon \cite{binci_first-principles_2021}. Crucially, in momentum space this takes the form $V_\mathrm{l-ph}= -\sqrt{N} F(t) \varphi_0(t)$, which means: (1) only the zone-center mode (phonon at the $\Gamma$-point) is driven and (2) it is driven with a force proportional to $\sqrt{N}$. The second point is the fundamental reason why macroscopically driven phonons are treated classically in most cases. Upon driving, the ground state evolves to a coherent state, which is nothing but the ground state displaced in phase space by an extensive amount. As such, the relative uncertainty
\begin{equation}\label{eq:relative_uncertainty}
    \sigma_\text{rel} = \frac{\sqrt{\expval{\varphi_0^2} - \expval{\varphi_0}^2}}{\abs{\expval{\varphi_0}}}
\end{equation}
vanishes for macroscopic systems like $1/\sqrt{N}$. The numerator in eq.~\eqref{eq:relative_uncertainty} is the square root of the second cumulant
\begin{equation}
    \expval{\varphi_0^2}_c := \expval{\varphi_0^2} - \expval{\varphi_0}^2 = \frac{\hbar}{2\omega_0}
\end{equation}
of the quantum generating functional.
It is noteworthy that its value is completely independent of linearly coupled external driving for a non-interacting phonon.\\

While classical treatments of nonlinear phononics are ubiquitous, we demonstrate in the following that even in the thermodynamic limit, a quantitative description of driven interacting phonons requires extending beyond this approximation. A quantum treatment is necessary for two primary reasons: (1) higher-order cumulants enter the equations of motion for the amplitude expectation values on equal footing with the coherent forcing, and (2) the drive endows these cumulants with independent dynamics, meaning they cannot simply be absorbed into renormalized parameters of the classical equations.\\
We use the prototypical $Q_\IR^2 Q_\R$ model of nonlinear phononics to illustrate these points. 
The microscopic action of the interacting two-phonon system across the Brillouin zone takes the general form
\begin{widetext}
\begin{equation}\label{eq:full_micro_Lagrangian}
\begin{split}
    S &= \sum_\q \int_{t}\left[\frac{1}{2}\left(\varphi_{\R,-\q} G_{0,\R,\q}^{-1}\varphi_{\R,\q} + \varphi_{\IR,-\q} G_{0,\IR,\q}^{-1}\varphi_{\IR,\q}\right) - \frac{1}{\sqrt{N}}\sum_{\qp} V_{\q,\qp}\; \varphi_{\R,-\q - \qp} \varphi_{\IR,\q} \varphi_{\IR,\qp}\right]\,,
\end{split}
\end{equation}
\end{widetext}
In this case, the inverse \emph{bare} propagators are $ G_{0,\R/\IR}^{-1}(\q) = - (\partial_t^2 + \omega_{\R/\IR,\q}^2)$.
Two significant simplifications can be applied to~\eqref{eq:full_micro_Lagrangian}: First, we approximate the phonon dispersions as constant, that is, $\omega_{\R/\IR,\q} \approx \omega_{\R/\IR}$, capturing the behavior of optical phonons. Second, the momentum-dependent vertex is also replaced by an effective constant $V_{\q,\qp} \approx \frac{\alpha}{2}$, which may be thought of as the average value across the Brillouin zone (BZ).\\
Using both simplifications, the system reduces to two effective Einstein phonons interacting locally in real-space according to the action
\small
\begin{equation}\label{eq:action}
    S = \sum_\textbf{r}\int_t \frac{1}{2}\left(Q_{\scriptscriptstyle \R} G_{\scriptscriptstyle 0,\R}^{-1}Q_{\scriptscriptstyle \R} + Q_{\scriptscriptstyle \IR} G_{\scriptscriptstyle 0, \IR}^{-1} Q_{\scriptscriptstyle \IR}  - \alpha Q_{\scriptscriptstyle \R} Q_{\scriptscriptstyle \IR}^2\right).
\end{equation}
\normalsize
The merit of starting from the realistic system~\eqref{eq:full_micro_Lagrangian} is that it clarifies that the zone-center modes do not just interact among themselves. They interact with an \emph{extensive} number of other phonons, that is, with all pairs of phonons at $\q$ and $-\q$ \footnote{For this extensivity, it is not necessary for the vertex $V_{\q,\q'}$ to be nonzero for all momenta. It suffices if $V_{\q,\q'}$ takes finite values on any (non-empty) open subset of the BZ.}.\\

\subsection{\label{sec:thermo_limit_scaling}  Fluctuations in the thermodynamic limit}

To derive corrections to the dynamics due to quantum and thermal fluctuations systematically, we use the 1-particle irreducible (1PI) effective action formalism \cite{berges_introduction_2004,rammer_quantum_2007,jona-lasinio_relativistic_1964,coleman_radiative_1973}. Consider the quantum partition function
\begin{equation}
    Z[J] = \int D\varphi \, e^{i\left(S[\varphi]-\varphi\cdot J \right)}\,,
\end{equation}
where, for notational compactness, we denote tuples containing all field components by variables without the respective indices, e.g. $\varphi = \begin{pmatrix}
    \varphi_\R & \varphi_\IR
\end{pmatrix}$ and use the interior product notation
\begin{equation}\label{eq:inner_product}
    \varphi \cdot J = \sum_{j\in \R,\IR} \sum_\q \int_t\, \varphi_j(\q,t) J_j(-\q,t)\,.
\end{equation}
Here $J = \begin{pmatrix}
   J_\R & J_\IR
\end{pmatrix}$ is an arbitrary external source field, which we eventually equate to the electric field driving such that
\begin{align}
    J_\IR(\q,t) =  -\sqrt{N} F(t)\delta_{\q,0} \quad \text{and} \quad J_\R(\q,t) \approx 0\,.
\end{align}
The Schwinger functional and 1PI effective action are defined as
\begin{align}
        W[J] &= i \ln{Z[J]}\quad\text{and}\\
        \Gamma[ \phi] &= \sup_J\left( \phi \cdot  J - W[J]\right)\,,
\end{align}
such that $\frac{\delta W[J]}{\delta J_i(t)} = \phi_i(t)$ and $\Gamma_i^{(1)}(t) \coloneqq \frac{\delta \Gamma[\phi]}{\delta \phi_i(t)}  = J_i(t)$. Here and below, we use $\phi = \expval{\varphi}$ for the field expectation value. 
The Dyson-Schwinger master equation generates the equations of motion of all n-point functions ($\Gamma^{(n)}$) via successive differentiation \cite{swanson_primer_2010},
\begin{equation}\label{eq:Dyson-Schwinger}
    \frac{\delta \Gamma [\phi]}{\delta \phi} = \frac{\delta S [\varphi]}{\delta \varphi}\Bigg\vert_{\varphi = iG\cdot \frac{\delta}{\delta \phi} + \phi}\,.
\end{equation}
Here, $G$ denotes the fully dressed propagator
\begin{equation}
    G_{jk}(1,2) = -i\langle \mathcal{T}\varphi_j(1) \varphi_k(2) \rangle_c
\end{equation} 
with ${j,k}\in \{\R,\IR\}$ and $\mathcal{T}$ denoting the contour ordering operator. For compactness, we abbreviate external arguments using integer superindices.
By evaluating \cref{eq:Dyson-Schwinger}, we directly obtain the field expectation value equations of motion:
\begin{widetext}
    \begin{subequations}\label{eq:field-eoms}
    \begin{align}
        J_\R(\q,t) &= G_{0,\R}\inv \PhR(\q,t) - \frac{\alpha}{2\sqrt{N}} \sum_{\q'}\left(\PhIR(\q' + \q,t)\PhIR(-\q',t) + i\, G_{\IR}(\q' + \q,t;-\q',t)\right)\\
    J_\IR(\q,t) &= G_{0,\IR}\inv \PhIR(\q,t) - \frac{\alpha}{\sqrt{N}} \sum_{\q'} \left( \PhIR(\q' + \q,t)\PhR(-\q',t) + i\, G_{\R\IR}(\q' + \q,t; - \q',t)\right) 
    \end{align} 
\end{subequations}
\end{widetext}
This is identical to the result found by taking the expectation value of the microscopic field EoM
\begin{equation}
    J_{\R/\IR}(\q,t) = \expval{\frac{\delta S}{\delta \varphi_{\R/\IR}(\q,t)}}\,.
\end{equation}
The EoM Eqs.~\eqref{eq:field-eoms} are distinct from the classical equations of motion ($\frac{\delta S[\varphi]}{\delta \varphi}=0$) owing to the presence of the propagator, as a direct consequence of not assuming the classical behavior $\expval{\varphi\varphi}=\expval{\varphi}^2$.\\
Because the external drive is assumed to be homogeneous ($\propto \delta_{\q,0}$) for optical light, translation invariance leaves only the zone-center mode equations nonzero. Furthermore, the zone-center mode equations of motion simplify using that $\phi(\q,t) = 0$ unless $\q = 0$, and because we approximated the action as momentum-independent, $\sum_\q G_\q(t,t) = N G(t,t)$. We drop the zero-momentum argument and use the shorthand $\phi(t)$ for the zone-center mode $\phi(\q = 0,t)$, which evolves according to the simplified EoM
\small
\begin{subequations}
\begin{align}
    0 &= G_{0,\R}\inv \PhR(t) - \frac{\alpha}{2\sqrt{N}} \Big(\PhIR(t)\PhIR(t) + iN\,  G_{\IR}(t,t)\Big) \label{eq:phi_R_EoM_simplified}\\
    \begin{split}
    0 &= G_{0,\IR}\inv \PhIR(t) - \frac{\alpha}{\sqrt{N}} \Big( \PhIR(t)\PhR(t) + iN\,G_{\R\IR}(t,t)\Big)\\
    &\quad + \sqrt{N}F(t)\label{eq:phi_IR_EoM_simplified}
    \end{split}
\end{align} 
\end{subequations}
\normalsize
As a consequence of the extensive drive $\sqrt{N} F(t)$, the field expectation values scale identically as $\phi \sim \mathcal{O}\left(\sqrt{N}\right)$. However, from Eq.~\eqref{eq:phi_R_EoM_simplified} and~\eqref{eq:phi_IR_EoM_simplified}, fluctuation effects can be seen to scale the same. Both terms in the interaction brackets are $\mathcal{O}\left(N\right)$. If we introduce a rescaled field $\bar \phi(t) = \phi(t)\slash \sqrt{N}$, all occurrences of $N$ disappear from its EoM. In conclusion, the relevance of quantum corrections in the amplitude expectation value equations of motion is independent of the system size $N$.\\

If the value of $G(t,t)$ were given by the static zero-point fluctuation $-iC_0$, its entire effect could be captured by renormalizing the model, e.g., the Raman amplitude would be shifted according to
\begin{equation}
    \bar \phi_\R \to \bar \phi_\R + \frac{\alpha C_0}{2\omega_\R^2}\quad \text{and}\quad \omega_\IR^2 \to \omega_\IR^2 - \frac{\alpha^2 C_0}{2\omega_\R^2}\,.
\end{equation}

The thus shifted amplitude evolves according to the classical EoM. In order for fluctuation effects to become truly relevant, they must be dynamical. \\
Applying the chain rule of functional differentiation on the defining equations of $\Gamma[\phi]$ and $W[J]$, one obtains the well-known relation
\begin{equation}\label{eq:Gamma2=G_inv}
    \Gamma^{(2)}_{ij}(1,2) := \frac{\delta^2\Gamma}{\delta \phi_i(1) \delta\phi_j(2)} = G^{-1}_{ij}(1,2) \,.
\end{equation}
Here, $\Gamma^{(2)}$ and $G$ are related by an inversion in both the external (time and space) and internal (phonon type) indices. From \cref{eq:Gamma2=G_inv}, we obtain the propagator in terms of the proper 2-point vertex $\Gamma^{(2)}$ via block matrix inversion. 
\begin{fmffile}{props_and_gamma2s_R}
\begin{subequations}
\begin{align}
\label{eq:propagators_in_terms_of_propagators}
    \begin{split}
    G_\R &= {\Gamma^{(2)}_\R}^{-1} - {\Gamma^{(2)}_\R}^{-1} \cdot \Gamma^{(2)}_{\R\IR} \cdot G_{\IR\R}\\
    \eqgraph{0ex}{0ex}{
    \begin{fmfgraph}(10,10)
        \fmfleft{i}
        \fmfright{o}
        \fmf{dbl_zigzag,width=1}{i,o}
    \end{fmfgraph}}
    &=\;
    \eqgraph{0ex}{0ex}{
    \begin{fmfgraph}(10,10)
        \fmfleft{i}
        \fmfright{o}
        \fmf{zigzag,width=2}{i,o}
    \end{fmfgraph}}
    -\;
    \eqgraph{0ex}{0ex}{
    \begin{fmfgraph*}(22,10)
        \fmfleft{i}
        \fmfright{o}
        \fmf{zigzag,width=2,tension=0.6}{i,v1}
        \fmf{dbl_plain,tension=0.75}{v1,v2}
        \fmf{dbl_zigzag,width=1}{v2,o}
        \fmfv{d.sh=circle,d.f=empty,d.si=10}{v1}
    \end{fmfgraph*}}
    \end{split}\\
    \begin{split}
    G_\IR &= {\Gamma^{(2)}_\IR}^{-1} - {\Gamma^{(2)}_\IR}^{-1} \cdot \Gamma^{(2)}_{\IR\R} \cdot G_{\R\IR} \\
    \eqgraph{0ex}{0ex}{
    \begin{fmfgraph}(10,10)
        \fmfleft{i}
        \fmfright{o}
        \fmf{dbl_plain,width=1}{i,o}
    \end{fmfgraph}}
    &=\;
    \eqgraph{0ex}{0ex}{
    \begin{fmfgraph}(10,10)
        \fmfleft{i}
        \fmfright{o}
        \fmf{plain,width=2}{i,o}
    \end{fmfgraph}}
    -\;
    \eqgraph{0ex}{0ex}{
    \begin{fmfgraph*}(22,10)
        \fmfleft{i}
        \fmfright{o}
        \fmf{plain,width=2,tension=0.6}{i,v1}
        \fmf{dbl_zigzag}{v1,v2}
        \fmf{dbl_plain,width=1}{v2,o}
        \fmfv{d.sh=circle,d.f=empty,d.si=10}{v1}
    \end{fmfgraph*}}
    \end{split}\\
     \begin{split}
    G_{\IR\R} &= -{\Gamma^{(2)}_\IR}^{-1} \cdot \Gamma^{(2)}_{\IR\R} \cdot G_\R\\
    \eqgraph{0ex}{0ex}{
    \begin{fmfgraph}(10,10)
        \fmfleft{i}
        \fmfright{o}
        \fmf{dbl_plain,width=1}{i,v1}
        \fmf{dbl_zigzag,width=1}{v1,o}
    \end{fmfgraph}}
    &=\; - \;
    \eqgraph{1ex}{0ex}{
    \begin{fmfgraph*}(20,10)
        \fmfleft{i}
        \fmfright{o}
        \fmf{plain,width=2}{i,v1}
        \fmf{dbl_zigzag,width=1}{v1,o}
        \fmfv{d.sh=circle,d.f=empty,d.si=10}{v1}
    \end{fmfgraph*}}
    \end{split}
\end{align}
\end{subequations}
\end{fmffile}
We use zigzags for the Raman-active phonons and plain lines for the IR-active phonons, bold lines for the inverse two-point functions ${\Gamma^{(2)}_{ij}}^{-1}$ and doubled lines for the full propagators $G_{ij}$. A comprehensive list of the Feynman diagram notations used here and in the following can be found in Appendix \ref{app: Feynman_rules}.
The "mixed" propagators $G_{\IR\R}, G_{\R\IR}$ as well as the oscillation between IR and R modes in the respective propagators are absent in the equilibrium state, where the action is symmetric under $\varphi_\IR \rightarrow -\varphi_\IR$. However, including the external source breaks this symmetry, so correlation functions with odd numbers of $\varphi_\IR$ fields are no longer forced to vanish. Therefore, an additional layer of dressing is added to the propagators in the presence of a symmetry-breaking source. The bare propagators are dressed by self-energy insertions (thick propagators in the diagrams). Additional resummation is required to capture the series of oscillations between the Raman and IR modes with intermediate propagation in each mode, as shown in \cref{eq:propagators_in_terms_of_propagators}.
The equations of motion of the 2-point functions (the Dyson equations) are obtained by differentiating~\eqref{eq:Dyson-Schwinger} once more. In general, the Dyson-Schwinger equation of the n-point function $\Gamma^{(n)}$ references at least the (n+1)-point function $\Gamma^{(n+1)}$, forming an infinite tower of coupled integro-differential equations. For simplicity, we consider quantum corrections only up to 2-point functions. Consequently, we approximate the 3-point vertex by its bare value, 
\begin{equation}
    \Gamma_{ijk}^{(3)}(1,2,3) = \begin{cases}
        -\alpha\,\delta_{123},\; \{i,j,k\} = \{\IR,\IR,\R\}\\
        0,\; \text{else}
    \end{cases} 
\end{equation}
and all higher-order vertices are assumed to vanish. Employing this approximation, we derive the Dyson equations for the 2-point functions:
\begin{widetext}
\begin{fmffile}{Selfenergies}
\fmfcmd{
    path quadrant, q[], otimes;
    quadrant = (0, 0) -- (0.5, 0) & quartercircle & (0, 0.5) -- (0, 0);
    for i=1 upto 4: q[i] = quadrant rotated (45 + 90*i); endfor
    otimes = q[1] & q[2] & q[3] & q[4] -- cycle;
}
\fmfwizard
\begin{subequations}\label{eq:Selfenergies}
\begin{gather}
\begin{split}\label{eq:Gamma_R}
    \frac{\delta^2 \Gamma}{\delta \PhR(1)\delta \PhR(2)}
    &\approx G_{0,\R}^{-1}  - i \frac{\alpha^2}{2} G_\IR(1,2) G_\IR(2,1)\\ 
    \eqgraph{0ex}{0ex}{
    \begin{fmfgraph}(10,10)
        \fmfleft{i}
        \fmfright{o}
        \fmf{zigzag,width=2}{i,o}
    \end{fmfgraph}}^{-1}
    \;&\approx \,
    \eqgraph{0ex}{0ex}{\begin{fmfgraph*}(10,10)
        \fmfleft{i}
        \fmfright{o}
        \fmf{zigzag,width=1}{i,o}
    \end{fmfgraph*}}^{-1}\; - \; \frac{i}{2} \,
    \eqgraph{0ex}{0ex}{\begin{fmfgraph*}(20,10)
        \fmfleft{i}
        \fmfright{o}
        \fmf{zigzag,width=1}{i,v1}
        \fmf{zigzag,width=1}{v2,o}
        \fmf{dbl_plain,left=1,tension=0.2}{v1,v2,v1}
        \fmfdot{v1,v2}
    \end{fmfgraph*}} 
    \end{split}\\
\begin{split}\label{eq:Gamma_IR}
    \frac{\delta^2 \Gamma}{\delta \PhIR(1)\delta \PhIR(2)}
    &\approx G_{0,\IR}^{-1}  - \alpha \bar \phi_\R(1) \delta_{1,2} - i \alpha^2 G_{\R\IR}(1,2) G_{\R\IR}(2,1) - i \alpha^2 G_{\R}(1,2) G_{\IR}(2,1)\\ 
    \eqgraph{0ex}{0ex}{
    \begin{fmfgraph}(10,10)
        \fmfleft{i}
        \fmfright{o}
        \fmf{plain,width=2}{i,o}
    \end{fmfgraph}}^{-1}
    \;&\approx\;
    \eqgraph{0ex}{0ex}{\begin{fmfgraph*}(10,10)
        \fmfleft{i}
        \fmfright{o}
        \fmf{plain,width=1}{i,o}
    \end{fmfgraph*}}^{-1}\;+\quad 
    \eqgraph{0ex}{0ex}{\begin{fmfgraph*}(10,10)
        \fmfleft{i}
        \fmfright{o}
        \fmftop{t}
        \fmf{plain,width=1}{i,v1,o}
        \fmf{zigzag,left=0,tension=0.}{v1,t}
        \fmfdot{v1}
        \fmfv{d.sh=otimes,d.f=empty,d.si=10,l.d=6,l.a=0,l=$\bar \phi_\R$}{t}
    \end{fmfgraph*}}\quad -\quad i\;
    \eqgraph{0ex}{0ex}{\begin{fmfgraph*}(20,10)
        \fmfleft{i}
        \fmfright{o}
        \fmftop{t}
        \fmfbottom{b}
        \fmf{plain,width=1}{i,v1}
        \fmf{plain,width=1}{v2,o}
        \fmf{dbl_plain,left=0.3,tension=0.2}{v1,t}
        \fmf{dbl_zigzag,left=0.3,tension=0.2}{t,v2}
        \fmf{dbl_plain,left=0.3,tension=0.2}{v2,b}
        \fmf{dbl_zigzag,left=0.3,tension=0.2}{b,v1}
        \fmfdot{v1,v2}
    \end{fmfgraph*}} \quad -\quad i\; 
    \eqgraph{0ex}{0ex}{\begin{fmfgraph*}(20,10)
        \fmfleft{i}
        \fmfright{o}
        \fmf{plain,width=1}{i,v1}
        \fmf{plain,width=1}{v2,o}
        \fmf{dbl_plain,left=0,tension=0.2}{v1,v2}
        \fmf{dbl_zigzag,left=1,tension=0.2}{v1,v2}
        \fmfdot{v1,v2}
    \end{fmfgraph*}} 
    \end{split}\\
\begin{split}\label{eq:Gamma_IRR}
    \frac{\delta}{\delta \phi_\R(2)}\frac{\delta \Gamma}{\delta \PhIR(1)} 
    &\approx -\alpha \bar \phi_\IR(2)\delta_{1,2} - i \alpha^2 G_{\R\IR}(1,2) G_\IR(2,1)\\ 
    \eqgraph{0ex}{0ex}{\begin{fmfgraph*}(10,10)
        \fmfleft{i}
        \fmfright{o}
        \fmf{plain,width=1}{i,v1}
        \fmf{zigzag,width=1}{v1,o}
        \fmfv{d.sh=circle,d.f=empty,d.si=10,l.d=0}{v1}
    \end{fmfgraph*}}\quad &\approx \quad
    \eqgraph{0ex}{0ex}{\begin{fmfgraph*}(10,10)
        \fmfleft{i}
        \fmfright{o}
        \fmftop{t}
        \fmf{plain,width=1}{i,v1}
        \fmf{zigzag,width=1}{v1,o}
        \fmf{plain,left=0,tension=0.}{v1,t}
        \fmfdot{v1}
        \fmfv{d.sh=otimes,d.f=empty,d.si=10,l.d=6,l.a=0,l=$\bar \phi_\IR$}{t}
    \end{fmfgraph*}}\quad - \quad i \;
    \eqgraph{0ex}{0ex}{\begin{fmfgraph*}(20,10)
        \fmfleft{i}
        \fmfright{o}
        \fmfbottom{b}
        \fmftop{t}
        \fmf{plain,width=1}{i,v1}
        \fmf{zigzag,width=1}{v2,o}
        \fmf{dbl_plain,left=0.3,tension=0.2}{v1,t,v2}
        \fmf{dbl_plain,left=0.3,tension=0.2}{v2,b}
        \fmf{dbl_zigzag,left=0.3,tension=0.2}{b,v1}
        \fmfdot{v1,v2}
    \end{fmfgraph*}} 
    \end{split}
    \end{gather}
\end{subequations}
\end{fmffile}
\end{widetext}

Because the external field does not couple directly to the propagators, their dynamics are determined only by the change of their self-energies, which depend on the driven amplitude expectation values through the field insertions marked by crossed circles in~\eqref{eq:Gamma_IR} and~\eqref{eq:Gamma_IRR}, and on the propagator itself.\\
The field insertions of all $\Sigma_\q(t,t')$ (even if there was momentum dependence) are given by the zone-center modes in the form
\begin{equation}
    \frac{\alpha}{\sqrt{N}}\phi(\q=0,t)\,\delta_{t,t'} = \alpha \bar \phi(t)\, \delta_{t,t'}\,,
\end{equation}
which is independent of the system size. Similarly, the fluctuation loops scale as
\begin{equation}
    \left(\frac{\alpha}{\sqrt{N}}\right)^2 \sum_\k G_{\q + \k}(t,t') G_\k(t',t) \sim \mathcal{O}(1)
\end{equation}
because $\mathcal{O}(N)$ virtual processes are summed. Hence, in the driven state, the mean-field insertions and fluctuation loops both scale intensively as $\mathcal{O}(1)$.\\
In summary, we have demonstrated that both the first (see~\eqref{eq:field-eoms}) and second cumulants of driven nonlinear phononic systems receive quantum corrections to their equations of motion. Crucially, these corrections do not decay with system size, nor can they be discarded as static. This observation is clearest in the effective Einstein phonon model, where we can exploit direct proportionalities like $\sum_\q G_\q = N\, G$. However, the same scaling argument holds for dispersive phonons with a momentum-dependent vertex, provided the vertex support occupies a subset of the Brillouin zone with nonzero measure, thereby coupling an extensive number of modes.\\

Despite these differences between interacting and harmonic phonons, for individual momentum modes, it still holds that
\begin{align}
    \phi(\q,t) &\sim \begin{cases}
        \mathcal{O}\left(\sqrt{N}\right) \q = 0\\
        0\quad \q \neq 0
    \end{cases}\\
    G_\q(t,t') &\sim \mathcal{O}(1)\,.
\end{align}
This hierarchy is relevant for experimental access to higher cumulants such as $G$.
By optical means, only $\q \approx 0$ modes are accessible, but simply measuring the quadratic expectation value $\expval{\varphi(0,t)\varphi(0,t')} = i G(t,t') + \phi(0,t) \phi(0,t')$ is challenging from a signal-to-noise perspective. An optical probe needs to be capable of measuring the connected correlator $G$ without measuring the macroscopic field amplitude. Pump-probe Raman scattering achieves this by separating the coherent and incoherent parts in terms of their scattering angles.

\section{Nonequilibrium Raman scattering on phonons}\label{sec:Nonequilibrium_Raman_theory}
Phononic light-matter coupling is derived by considering the energy of the system in an external electric field \cite{merlin_generating_1997,cheng_coherent_2013}, that is
\begin{equation}\label{eq:light-matter-coupling}
    H_\text{lmc} \approx - \vec \mu_0(Q) \cdot \vec{E} - \frac{1}{2}\chi_{ij}(E,Q) E_i E_j
\end{equation}
up to quadratic order in the electric field $\vec E$. Here, $\vec \mu = \vec \mu_0 + \chi \cdot \vec E$ denotes the electric dipole moment, which can depend on an external electric field via the polarizability $\chi$ and on the material modes, particularly on phonons, $Q$. In centrosymmetric materials, $\vec \mu_0(Q)$ cannot depend linearly on inversion-symmetric phonon amplitudes. The leading light-matter interaction for these modes is therefore Raman activity via the Raman tensor $R_{ij} = \partial\chi_{ij}/\partial Q$, which encodes how strongly the phonon modifies the electric polarizability.
The photon flux of spontaneous Raman scattering for a given outgoing mode with frequency $\omega_\out$ is proportional to the generation rate of that photon mode \cite{werner_nonequilibrium_2023}
\begin{equation}
    \frac{d\sigma(t)}{d\Omega d\omega} \propto \frac{d}{dt}\expval{ n_\out(t)}\,,
\end{equation}
where $n_\out = a^\dagger_\out a_\out$ is the number operator of the out-mode (characterized by a tuple of polarization and wave vector).
We approximate the light-phonon interaction $E_i R_{ij} E_j$ as a perturbation; that is, we assume the probe pulse is sufficiently weak not to cause significant excitations itself.
In a calculation analogous to that for nonequilibrium electronic Raman scattering \cite{shvaika_interpreting_2018,wang_theory_2018,werner_nonequilibrium_2023}, we find for the nonequilibrium nonresonant phononic Raman scattering process to the lowest order in perturbation theory:
\begin{widetext}
\begin{equation}\label{eq:nonequilibrium_Raman_response}
    \expval{ n_\out(t)} \propto \abs{E_\mathrm{in} \epsilon^\text{in}_i R_{ij} \epsilon^\text{out}_j }^2 \int_{-\infty}^t dt_+ \int_{-\infty}^t dt_-\, s(t_+) s(t_-) e^{i(\omega_\out - \omega_\text{in})(t_+-t_-)} \expval{\varphi(t_-) \varphi(t_+)}
\end{equation}
\end{widetext}
where the term in front of the integrals captures the light-matter coupling strength of the phonon as well as the polarization dependence, and $s(t)$ is the envelope of the probe beam, which we have assumed to be of the form 
\begin{equation}
    \vec E_\text{in}(t) = E_\mathrm{in} \hat\epsilon_\text{in} s(t) \sin(\omega_\text{in} t)\,.
\end{equation}
Importantly, we assumed that the Raman tensor $R_{ij}$ in~\eqref{eq:nonequilibrium_Raman_response} may depend on the probe frequency, but remains fixed at its equilibrium functional value throughout the nonequilibrium evolution. This approximation neglects changes in the Raman matrix elements induced by the pump or by lattice motion, allowing us to evaluate the Raman response from the phonon dynamics without explicitly computing the nonequilibrium electronic contribution to the Raman tensor. While this greatly simplifies calculations, it is difficult to estimate the error of this approximation systematically. In particular, it can be highly material- and mode-specific \cite{zhu_dynamical_2019, henstridge_nonlocal_2022}. Corrections can be estimated by expanding the electronic polarizability tensor about equilibrium in the driven phonon coordinates and electric field, with higher-order derivatives obtained, for example, using time-dependent density functional perturbation theory (TDDFPT). However, such an equilibrium-based expansion still does not account for pump-induced changes in electronic occupations or the full history dependence of the nonequilibrium electronic response.

The measured intensity is proportional to the integrated emission over the probe duration, i.e., taking the limit $t\to\infty$ in eq.~\eqref{eq:nonequilibrium_Raman_response}. The probe envelope width then sets the time resolution.
Up to a polarization-dependent proportionality constant, the relevant observable that is probed by spontaneous Raman scattering on phonons is the \emph{full} greater correlator,
\begin{equation}\label{eq:full_greater_propagator}
    C^>(t,t'):= \expval{\varphi(t_-)\varphi(t'_+)}\,,
\end{equation}
where the $\pm$ indices indicate the ordering on the Keldysh contour. In \cref{eq:full_greater_propagator,eq:nonequilibrium_Raman_response}, the Raman probe beam does not enter the system dynamics. Therefore, we do not capture self-excitation effects, which become relevant for probe beams with a large field strength. However, we capture the pump beam nonperturbatively in the system's time evolution, with the pump field as an external source.
The full correlator can be expressed in terms of connected and disconnected parts,
\begin{equation}
    C^>(t,t') = i G^>(t,t') + \phi(t) \phi(t')\,.
\end{equation}
The connected part, given by the propagator $G^>$, captures the fluctuation-induced contribution. Because the fluctuations in the probed volume have no spatial coherence, the material acts as an array of randomly phased antennas oriented along the induced dipole moment and emits scattered photons in almost all directions. In contrast, the disconnected part captures a coherent oscillation absent in equilibrium Raman scattering, because without a prior pump, the coherent phonon amplitude $\phi(t) = \expval{\varphi(t)}$ is constant. Because the pump-induced oscillation is spatially coherent over the area illuminated by the pump beam, this component induces spatially coherent oscillating dipole moments akin to a phased array of antennas (see Appendix \ref{sec:coherent_Raman_derivation}).\\
 The two coherent sidebands radiate at surface-normal angles $\alpha_\pm$ distinct from the direct reflection angle (see Fig. \ref{fig:Raman_sketch}). These can be obtained by exploiting momentum conservation of light in the surface-parallel direction, which implies that
\begin{equation}
   \frac{\Omega_\text{in}}{c}\sin(\alpha_\text{in}) \pm q_\parallel = \frac{\Omega_\text{in}\pm \Omega_\text{ph}}{c}\sin(\alpha_\pm) \,,
\end{equation}
where $q_\parallel$ is the in-plane component of the coherent phonon wave vector. For a normal-incident pump beam, the surface-parallel component $q_\parallel = 2\frac{\Omega_\text{pump}}{c}\sin(\alpha_\text{pump})$ vanishes. Given a $515\,$nm probe and incidence angles $\alpha_\text{in}$ between 0° and 45°, as well as a $6.2\, \mu $m pump beam at normal incidence, the sidebands are emitted at angles $15\%$ to $30\%$ distinct from the incidence angle. 
This allows for targeted measurement of the sidebands without the fundamental background. The deflection angles of the sidebands in this scenario, specific to the \eg phonon in graphite, are particularly pronounced because the phonon's coherent oscillation frequency is close to the probe frequency (approximately 1/6). For softer optical phonons, the effect diminishes but may be enhanced by choosing shallower probe incidence angles.\\
In addition to its anisotropy, the coherent scattering contribution is further distinguished from the incoherent contribution by the ratio between its "Stokes" and "anti-Stokes" amplitudes. For electronically mediated Raman scattering with $\Omega_\text{in}\gg\Omega_\text{ph}$
the two sidebands have approximately equal intensity. Furthermore, since the coherent sidebands track the oscillatory components of the mean displacement, whereas spontaneous Raman features probe the driven connected correlation function, their spectral positions need not coincide.\\


 \section{Semiclassical approximation}\label{sec:Semiclassical}
\begin{figure}
    \centering
    \includegraphics[width=0.99\linewidth]{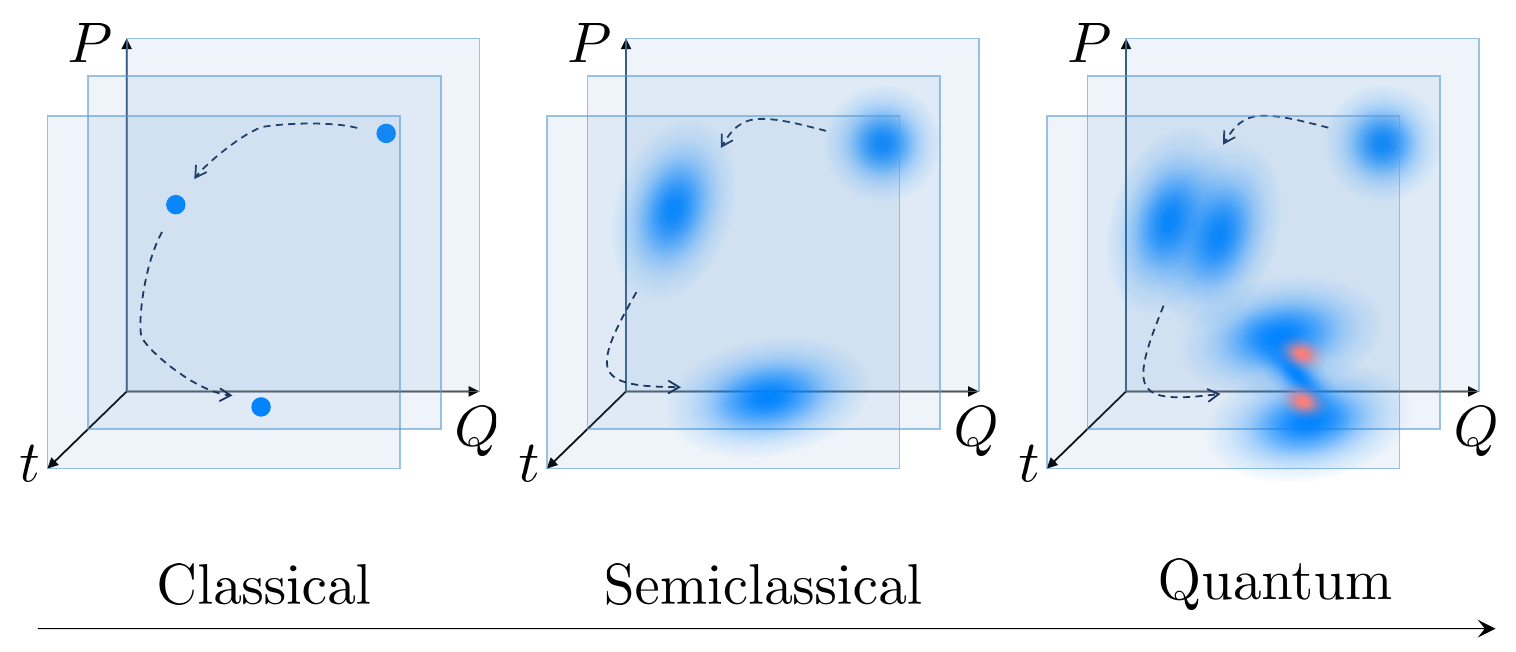}
    \caption{Illustration of the Wigner function dynamics in phase space for the classical, semiclassical, and quantum theory. In classical mechanics, the system occupies a single point in phase space with probability of one. The only dynamics is the movement of that point in phase space. In the Gaussian state approximation, the variance of the distribution is an additional degree of freedom that evolves in time. Finally, for full quantum time evolution, the Wigner distribution may take negative values (red) and is not restricted to a Gaussian shape.}
    \label{fig:semiclassical}
\end{figure}
For many systems, the phonon-phonon coupling parameters are small enough to use (self-consistent) perturbation theory. In this case, we anticipate that the loop corrections in the self-energies~\eqref{eq:Selfenergies} are subdominant compared to the mean-field insertions based on power-counting of the coupling. We can therefore make a semiclassical approximation that includes fluctuations, but accounts only for their interactions with the mean fields, and not among themselves.
For the cubic theory~\eqref{eq:action}, the equations of motion in this approximation are obtained by the cumulant expansion to second order \cite{kubo_generalized_1962} (Gaussian state approximation), or by using the 1-loop approximation of the two-particle irreducible (2PI) effective action (see Appendix \ref{sec:2PI_derivation}). The latter guarantees that the approximation is energy-conserving or energy-balancing in the driven open-system scenario~\cite{baym_conservation_1961,baym_self-consistent_1962,berges_controlled_2002}.\\
The time-locality of the self-energies in this approximation decouples the evolution of time-local observables, $\phi(t)$ and $G(t,t)$, from that of $G(t,t')$ for $t' \neq t$. The equations of motion for the time-local observables are
\begin{subequations}\label{eq:time_local_EoMs}
    \begin{align}
    \begin{split}
        \Ddot{\bar\phi}_\R(t) &= - \omega_\R^2 \bar\phi_\R(t) - \gamma_\R \dot{\bar\phi}_\R(t)\\
        &\quad - \frac{\alpha}{2}\left(\bar\phi_\IR(t)^2 + U_{\IR\IR}(t)\right)
    \end{split}\\
    \begin{split}
        \Ddot{\bar\phi}_\IR(t) &= F(t) - \omega_\IR^2 \bar\phi_\IR(t) - \gamma_\IR \dot{ \bar\phi}_\IR(t)\\
        &\quad - \alpha\left(\bar\phi_\IR(t)\bar\phi_\R(t) + U_{\IR\R}(t)\right)
    \end{split}\\
    \dot U(t) &= V^T(t) + V(t)\,,\label{eq:dU}\\
    \dot V(t) &= W(t) - U(t) \cdot M_\phi^2(t) - V \cdot \Gamma\,,\label{eq:dV}\\
    \begin{split}
        \dot W(t) &= - M_\phi^2(t) \cdot V(t) - V^T(t) \cdot M_\phi^2(t)\\
    &\quad - \Gamma \cdot W(t) - W(t)\cdot \Gamma + D\,.\label{eq:dW}
    \end{split}
    \end{align}
\end{subequations}
where we identified the symmetric Keldysh component of the propagator $G_{ij}^S(1,2) = -\frac{i}{2}\expval{\{\varphi_i(1),\,\varphi_j(2)\}}_c$ and defined the abbreviations
\begin{subequations}
    \begin{align}
    U(t) &:= iG^S(t,t)\,,\\
    V(t) &:= i\partial_{t'} G^S(t,t')\Big\vert_{t'=t}\,,\\
    W(t) &:= i\partial_{t}\partial_{t'} G^S(t,t')\Big\vert_{t'=t}\,,
\end{align}
\end{subequations}
as well as the mean-field potential curvature
\begin{equation}
    M_\phi^2(t) := \begin{pmatrix}
        \omega_\R^2 & \alpha \bar\phi_\IR(t) \\
        \alpha \bar\phi_\IR(t) & \omega_\IR^2 + \alpha \bar \phi_\R(t)
    \end{pmatrix}\,.
\end{equation}
To include the effect of external degrees of freedom, in particular acoustic phonons and electrons, we have added a phenomenological embedding self-energy 
\begin{equation}
    S \to S_\mathrm{open} = S - \frac{1}{2} \int_{1,2}\varphi_i(1) \Sigma_{\mathrm{em},ij}(1,2) \varphi_j(2)
\end{equation}
with the embedding rate function $\Gamma(\omega) = -2\Im \Sigma^\ret_\mathrm{em} (\omega)$ given by
\begin{equation}\label{eq:regulated_rate_function}
     \Gamma(\omega) = 2\omega\begin{pmatrix}
\gamma_\R  & 0 \\
0 & \gamma_\IR  
\end{pmatrix} \equiv 2\omega\, \Gamma\,,
\end{equation}
thus generating ohmic damping with the damping rates $\gamma_\R$ and $\gamma_\IR$ for the \eg and \eu mode respectively \cite{weiss_quantum_2021}. In the Kadanoff-Baym equation (KBE) of the symmetric correlator from which Eqs. (\ref{eq:dU}-\ref{eq:dW}) are derived, the embedding contributes a noise kernel $\Sigma_\emb^S(t,t')$. In the Markov approximation we find $\Sigma^S_\emb(\omega) = -iD$ with $D = \mathrm{diag}\left(\omega_\R\gamma_\R,\,\omega_\IR\gamma_\IR\right)$.

Next we consider the non-local observables. Because the time-diagonal is already known, we obtain $G^>(t,t')$ by solving 
\begin{equation}\label{eq:greater_KBE}
    \partial_t^2 G^>(t,t') = -  M_\phi^2(t) \cdot G^>(t,t') - \Gamma \cdot \partial_t G^>(t,t')  
\end{equation}
for $t>t'$ starting from $t=t'$ for each $t'$ with the initial conditions
\begin{equation}
    iG^>(t,t) = U(t)\,,\quad i\partial_t G^>(t,t')\Big \vert_{t'=t} = V^T -\frac{i}{2}\mathcal{1}\,,
\end{equation}
which follows from $G^> = G^S + \frac{1}{2}G^A$ and using that $G^A(t,t) = 0$ by symmetry and $\partial_tG^A(t,t) = -1$ as a direct consequence of the canonical commutation relations. We present a more detailed derivation of the semiclassical equations of motion in Appendix \ref{sec:2PI_derivation}.\\

As a benchmark for the semiclassical approximation, we use the fully quantum time evolution for the density matrix $\rho$ according to the Lindblad equation
 \begin{equation}
     \partial_t \hat\rho(t) = - i [\hat H, \hat\rho] + \sum_i \gamma_i \left(\hat L_i \hat \rho \hat L_i^\dagger - \frac{1}{2}\{\hat L_i^\dagger \hat L_i, \hat \rho\}\right)\,,
 \end{equation}
 with Lindblad jump operators 
 \begin{subequations}
 \begin{align}
     \hat L^\mathrm{loss}_\R &= \sqrt{\frac{1}{2}\left(\coth{\frac{\omega_\R}{2T}}+1\right)} \hat b_\R\\
     \hat L^\mathrm{gain}_\R &= \sqrt{\frac{1}{2}\left(\coth{\frac{\omega_\R}{2T}}-1\right)}\hat b^\dagger_\R
 \end{align}
 \end{subequations}
 and analogously for the IR-mode, where $\hat b^{(\dagger)}_i$ are the ladder operators of mode $i$.
To calculate the Raman signal, the greater correlator  $C^>(t_1,t_2) = \expval{\hat Q^H_\R(t_1) \hat Q^H_\R(t_2)}$ must be computed, where the superscript $H$ denotes the Heisenberg-picture operator. The Lindblad time evolution yields this two-time correlation function via the quantum regression theorem (QRT) \cite{gardiner_quantum_2004}. For $t_1\geq t_2$:
\begin{equation}
    C^>(t_1,t_2) = \tr\left( \hat Q_\R\, V(t_1,t_2)\{ \hat Q_\R\, V(t_2,t_0)\{ \hat\rho_0\}\}\right)
\end{equation}
That is, we evolve the initial state $\hat\rho_0$ to $t_2$ and then evolve the operator $\hat X \hat\rho(t_2)$ to $t_1$, where $\hat X$ is measured. Here, $V(t_a, t_b)$ denotes the Lindblad time-evolution superoperator from $t_b$ to $t_a$. The matrix entries for $t_1 < t_2$ are obtained by hermiticity, that is $C^>(t_1,t_2) = C^>(t_2,t_1)^*$. Both the QRT and the Lindblad-equation formulation assume Markovian dynamics. Given the damping parameters $\gamma \approx 1 \, \mathrm{meV}$ and the effective memory decay rate $1/\tau_\mathrm{mem} \sim \sqrt{(2 \pi T)^2 + \omega_0^2} \approx 250 \, \mathrm{meV}$ this is an excellent approximation. The Lindblad equation is solved using the Julia package \code{QuantumToolbox.jl} \cite{mercurio_quantumtoolboxjl_2025}.\\

The Lindblad computation's time complexity is $\mathcal{O}\left(N_t^2 \times (\dim \mathcal{H})^2\right)$, where $N_t$ is the number of time steps, assuming a sparse Lindblad superoperator. It is only feasible using the Einstein phonon approximation in which the Hilbert space dimension is
  \begin{equation}
      \dim \mathcal{H} = \left(N^\mathrm{max}_\R + 1\right) \times \left( N^\mathrm{max}_\IR + 1\right)
  \end{equation}
  with $N^\mathrm{max}_{\R/\IR}$ the occupation numbers after which the infinite bosonic Hilbert space is truncated for the Raman/IR mode, respectively.\\
On the other hand, the semiclassical Eq.~\eqref{eq:time_local_EoMs} and Eq.~\eqref{eq:greater_KBE} can be solved in $\mathcal{O}(N_t)$ and $\mathcal{O}(N_t^2)$ time complexity, respectively. The order reduction compared to the cubic scaling often associated with Keldysh field theory stems from the self-energy's time locality. Unlike the Lindblad approach, the semiclassical theory is independent of the Hilbert space dimension and therefore generally much cheaper and easier to extend to momentum-dependent models.\\

\begin{figure}
    \centering
    \includegraphics[width=0.99\linewidth]{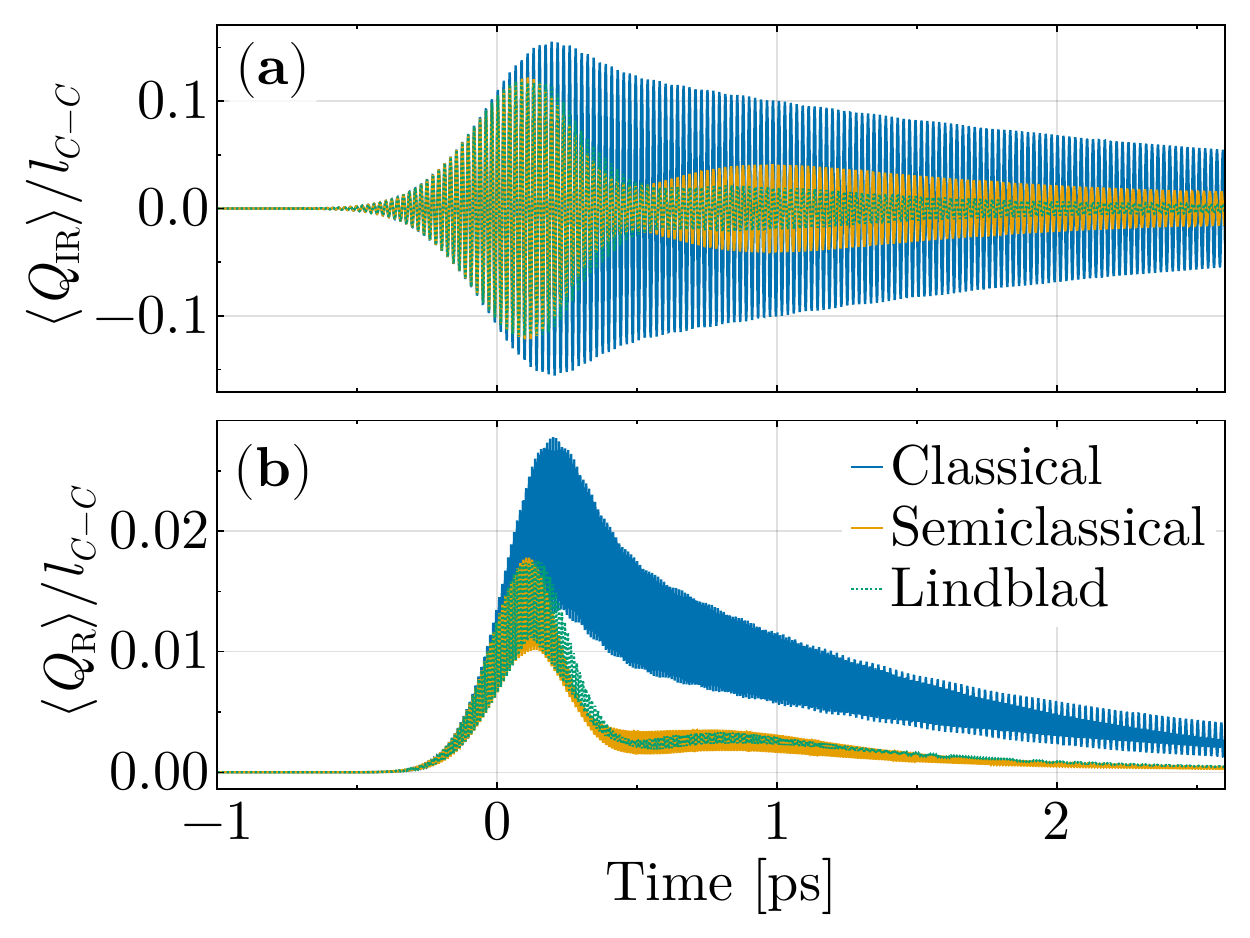}
    \caption{Comparison between the evolutions of the field amplitude expectation value of the IR-active (a) and Raman-active mode (b) in the classical, semiclassical, and Lindblad approximations upon a $500\, \mathrm{fs}$ in-resonance pulse with $F_\mathrm{max} = 0.1 \,\omega_\IR^{3/2}$ in the cubic theory of~\eqref{eq:action}. The model parameters are those of graphite, i.e., $\omega_\R = 196.1\, \mathrm{meV}$, $\omega_\IR = 196.9\, \mathrm{meV}$, $\gamma_\R = 1.5\, \mathrm{meV}$, $\gamma_\IR = 0.5\, \mathrm{meV}$, and $\bar \alpha = -0.1$. The phonon amplitudes are shown in units of the carbon-carbon bond length $l_{\mathrm{C-C}} = 1.42 \,\mathrm{\AA}$.}
    \label{fig:amplitude_evolution}
\end{figure}

To illustrate the role of fluctuation corrections to the evolution of the amplitude expectation values, we compare their evolution upon a short but very intense pump of the IR-mode using the classical, semiclassical, and Lindblad approximations, as shown in Fig.~\ref{fig:amplitude_evolution}. For this simulation, we used a simplified version of the multilayer graphene model of the next section. It hosts two nearly degenerate, very sharp modes, the IR-active \eu mode and the Raman-active \eg mode. However, we consider the simple interaction model~\eqref{eq:action} discussed so far. 
Perturbation theory in $\alpha$ provides an estimate for the relative change in the Raman resonance frequency for a given IR-mode amplitude $A$:
\begin{equation}
    \frac{\Delta \omega_\R}{\omega_\R} \approx -\frac{\alpha^2 A^2}{4\omega_\R^2(4\omega_\IR^2 - \omega_\R^2)}\,.
\end{equation}
For the simplified multilayer graphene system with $\omega_\R \approx \omega_0 \approx \omega_\IR$ and $\alpha = \bar{\alpha}\, \omega_0^{5/2}$, this simplifies to $\Delta \omega_\R/\omega_0 = - \bar{\alpha}^2 \bar{A}^2/12$, where we have identified the dimensionless amplitude $\bar A = \sqrt{\omega_0}A$. With $\bar\alpha=-0.1$, we can estimate the dimensionless IR-mode amplitude for a 1\% change in the Raman resonance frequency as $\bar A = \sqrt{12}$. Using the normal-mode mass $M = 12\,$u, we find this corresponds to an atomic displacement of $\abs{X} = \bar A /\sqrt{M\omega_0} \simeq 0.14\, \mathrm{\AA}$, about $10\%$ of the carbon-carbon bond length in graphene, which is around the upper limit of non-destructive pumping. We choose a pump field strength such that approximately $10\%$ displacement is reached at the maximum. Given the oscillator strength of $0.34 \, \mathrm{\AA}\,\mathrm{meV}^{1/2}$ (see Appendix \ref{sec:field_mediated_couplings}), we use a peak electric field strength of $81\,\mathrm{MV/cm}$ resulting in a pump fluence of $3.6\, \mathrm{J/cm}^2$ for the $0.5\,\mathrm{ps}$ FWHM pulse.
The Raman mode rectification reaches $\approx 1.4\%$ of the bond length in the Lindblad simulation. The amplitudes reached in the semiclassical and Lindblad simulations agree during the pump but disagree moderately in the decay phase beginning after $0.5\, \mathrm{ps}$. In contrast, the classical approximation significantly overestimates the amplitudes for most of the evolution. We attribute this to the (lack of) damping that the coherent amplitude experiences due to the second moments in the semiclassical and Lindblad simulations, that is absent in the classical approximation.
\\
\begin{figure}
     \centering
     \includegraphics[width=0.99\linewidth]{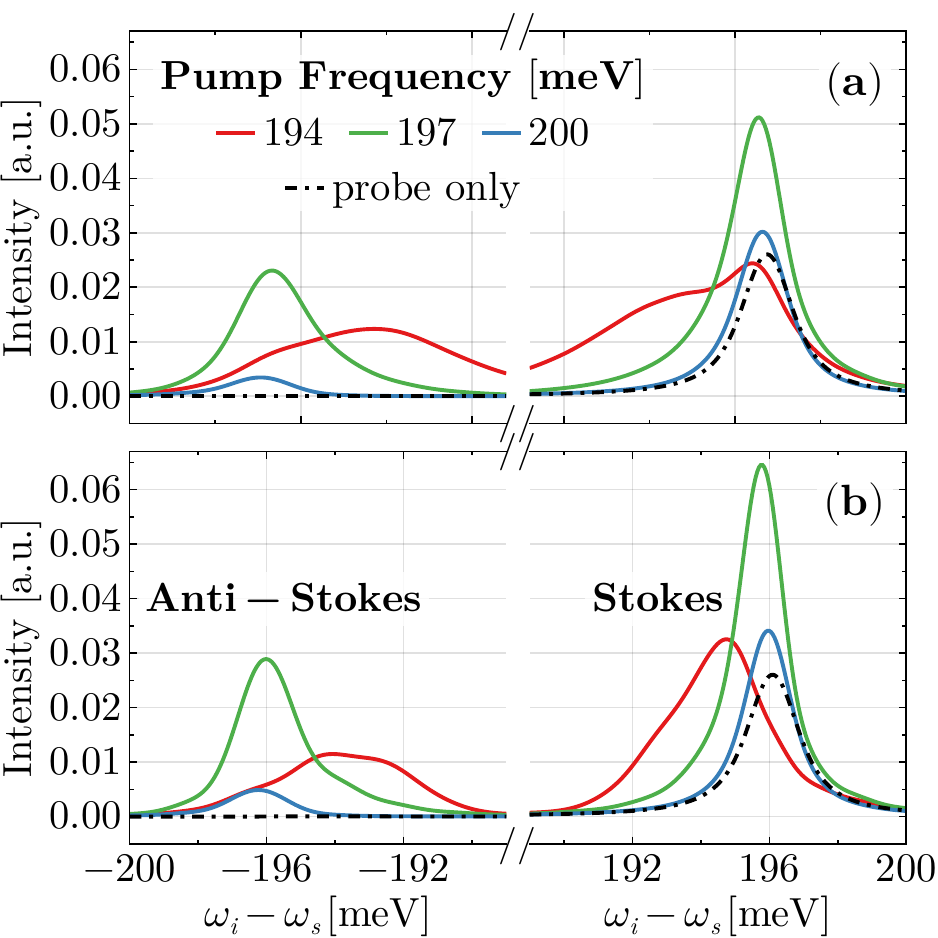}
     \caption{Nonequilibrium Raman spectrum for pump frequencies of $194,\, 197$ and $200\, \mathrm{meV}$, corresponding to red-detuned, in-resonance and blue-detuned driving, shown as red, green and blue lines, respectively. (a) Using the Lindblad formalism. (b) Using the semiclassical theory.  All pump beams have a FWHM of $0.5\,\mathrm{ps}$ and $F_\mathrm{max} = 0.1\, \omega_\IR^{3/2}$. The probe envelope is $4\, \mathrm{ps}$ wide and centered on the pump (zero time delay). The probe-only equilibrium Raman spectrum is shown as a dash-dotted line for comparison. The model parameters are those of graphite, i.e., $\omega_\R = 196.1\, \mathrm{meV}$, $\omega_\IR = 196.9\, \mathrm{meV}$, $\gamma_\R = 1.5\, \mathrm{meV}$, $\gamma_\IR = 0.5\, \mathrm{meV}$, and $\bar \alpha = -0.1$.}
     \label{fig:cubic_Raman_spectrum}
 \end{figure}
To compare the nonequilibrium fluctuation spectra obtained in the semiclassical approximation with the quantum theory, we simulate the nonequilibrium Raman signals due to $500 \,\mathrm{fs}$ pump pulses with $\hbar \Omega_\mathrm{drive} = 194\,, \quad 197\,,\quad \text{and} \quad 200 \,\mathrm{meV}$ using a perturbative $4\,\mathrm{ps}$ probe pulse in all cases. Mode softening occurs for all driving frequencies but is most pronounced for the red-detuned pump. For the in-resonance pulse, the Stokes signal doubles in intensity compared to equilibrium (dashed line), corresponding to an average occupation of one per unit cell over the probe duration and therefore an equivalent temperature of $3300\, \mathrm{K}$. However, the transient intensity does not arise from heating. In fact, the semiclassical approximation has no scattering mechanism that would allow thermalization except for dissipation into the zero-temperature environment, yet it shows an even larger transient signal than the Lindblad result. As with the amplitudes, we attribute the semiclassical approximation's higher intensity to the fact that some pump energy dissipates into higher moments in the Lindblad theory but cannot in the semiclassical approximation. Over the entire time evolution, the total energy pumped into the system differs by less than $6\%$ between the semiclassical and Lindblad simulations. The maximum energy deposited into the system for the in-resonance pump is $2.3\,\mathrm{eV}$ per unit cell, reached after approximately $190\,\mathrm{fs}$.\\
We conclude that, although much smaller modifications than presented in Fig.~\ref{fig:cubic_Raman_spectrum} are experimentally detectable, the pump fluence that is necessary to produce a measurable transient modification of the fluctuation spectrum of the \eg mode in the simplified graphite system is generally large, that is, on the order of J/cm$^2$. 
In general the necessary pump intensity for the quadratic-linear interaction scales as $1/\bar\alpha^2$, hence a gentler pump protocol suffices in materials with larger relative anharmonicity than our model's.

\section{Application to Multilayer Graphene}\label{sec:graphite}
Multilayer graphene in AB-stacking (HOPG) is an excellent system for investigating quantum effects in nonlinear phononics because its optical phonons have exceptionally high energies, and their thermal occupation is negligible at room temperature. This choice is further motivated by recent experimental results for this material \cite{buchenau_optically_2023}.\\
\begin{figure}\label{fig:phonon_sketch}
    \centering
    \includegraphics[width=0.9\linewidth]{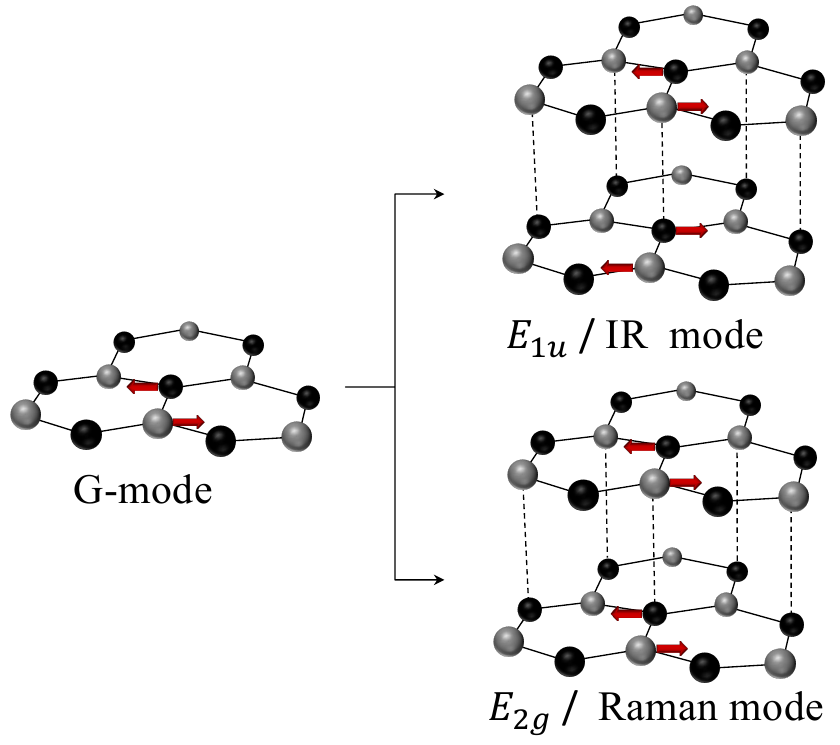}
    \caption{Illustration of the zone-center optical \eg and \eu phonons discussed in the main text. The G-mode in monolayer graphene becomes a Davydov doublet in the AB-stacked multilayer system. At the zone center, the direction of the red arrows describing the mode's polarization (drawn as transverse) is arbitrary, provided that it is in-plane and the two arrows belonging to the same unit cell point in opposite directions (making the mode optical). Figure inspired by \cite{nemanich_infrared_1977} and \cite{ferrari_raman_2013}.}
    \label{fig:phonon_sketch}
\end{figure}
The G-mode of graphene (see Fig.~\ref{fig:phonon_sketch}) is an unusually high-energy in-plane optical \eg mode consisting of the in-plane longitudinal optical (iLO) and in-plane transverse optical (iTO) phonon branches, which are degenerate at the Brillouin zone center \cite{ferrari_raman_2013}. Interlayer coupling in multilayer graphene/graphite yields a Davydov doublet of the symmetric \eg (sometimes denoted E$_{2g_2}$ in distinction to the much softer shear mode that also belongs to \eg \cite{cardona_light_1982})  and antisymmetric \eu stacking of the planes' G modes \cite{nemanich_infrared_1977}, as illustrated in Fig.~\ref{fig:phonon_sketch}.  In equilibrium, the IR-active \eu mode and the Raman-active \eg mode are nearly degenerate, owing to the weakness of the van der Waals interlayer coupling, with eigenenergies of approximately $196.9$ meV and $196.1$ meV, respectively \cite{ferrari_raman_2013,giura_temperature_2012,wirtz_phonon_2004,bonini_phonon_2007,nemanich_infrared_1977}. However, their linewidths at $300\,\mathrm{K}$ are quite distinct with $\gamma_\IR \approx 0.5 \,\mathrm{meV}$ and $\gamma_\R \approx 1.5\,\mathrm{meV}$ \cite{bonini_phonon_2007,giura_temperature_2012}.\\

Recently, it was found that driving the \eu phonon via a mid-IR pump produces a significant transient enhancement of the \eg Raman spectrum in multilayer but not in single-layer graphene. Furthermore the spectral signature of the \eg phonon is modulated analogously to the so-called optical spring effect known from cavity optomechanics \cite{aspelmeyer_cavity_2014}; which is to say, the \eg peak position shifts in the direction of the pump detuning when the pump frequency is close to the resonance frequency of the \eu mode, i.e., the \eg mode hardens for blue-detuned pumping and softens for red-detuned pumping of the \eu mode \cite{buchenau_optically_2023}. 
The fact that the pump affects the Raman-active phonon only when it is nearly resonant with the IR-active phonon and that the effect is absent in single-layer graphene suggests that \eu phonon activation is responsible for the effect \cite{buchenau_optically_2023}. Previously, the pump-induced peak shift was interpreted as an out-of-equilibrium form of level repulsion. However, the mechanism underlying this phenomenon remains unclear. The model proposed together with the experimental finding was an anharmonic coupling, $\frac{\alpha}{2} Q_\R Q_\IR^2$, between the two modes described by the classical equations of motion \cite{buchenau_optically_2023}
\begin{equation}\label{eq:classical_EOM}
\begin{split}
    \left(\partial_t^2 + \gamma_0\partial_t + \omega_0^2\right) Q_\IR(t) &= \alpha Q_\R(t) Q_\IR(t) + F(t)\\
    \left(\partial_t^2 + \gamma_0\partial_t + \omega_0^2\right) Q_\R(t) &= \frac{\alpha}{2} Q_\IR^2(t)
\end{split}
\end{equation}
where $Q_{\R/\IR}(t)$ is an effective normal mode amplitude of the \eg/\eu phonons respectively, and $F(t)$ is the driving. The interaction $Q_\R Q_\IR^2$, is a simplified proxy for the fully $D_{6h}$-invariant cubic coupling, $2u_xu_yg_x + g_y(u_x^2 - u_y^2)$, where $u_i$ and $g_i$ for $i\in x,y$ denote the two polarizations of the \eu and \eg modes respectively. The laser excites a superposition of the \eu polarizations that is parallel to the electric field. The distinction between polarizations is ignored due to their degeneracy at $\Gamma$. Using a frozen phonon calculation (appendix \ref{sec:frozen_phonons}), we do indeed find a significant coupling strength of $\bar \alpha \approx -0.1$ (in units of $\omega_\IR$) for the cubic interaction of the two zone-center modes (see Table \ref{tab:DFT_couplings}). 

However, as discussed in \ref{sec:Nonequilibrium_Raman_theory}, spontaneous Raman scattering measures the fluctuation spectrum, not the coherently driven amplitude. We have seen in the previous section that using the originally proposed model, a transient doubling of the Stokes side intensity as observed in the experiment requires a drastically larger pump fluence of $3.6\,\mathrm{J/cm}^2$ compared to the reported $2\, \mathrm{mJ/cm}^2$. To illustrate the point further, let us consider the energy per unit cell $\Delta \epsilon(t) = (\langle\hat H_\mathrm{ph}\rangle(t) - E_\mathrm{gs})/N$ that is pumped into the system. Here $\hat H_\mathrm{ph}$ denotes the phonon Hamiltonian (i.e., without the external field), and $E_\mathrm{gs}$ is the ground-state energy. Using second-order time-dependent perturbation theory, we find that this energy is bounded from above by
\begin{equation}\label{eq:absorption_bound}
    \Delta \epsilon(t) \leq \frac{\pi \abs{d_\IR}^2 (eE_0)^2 \sigma^2}{4}
\end{equation}
for a many-cycle Gaussian electric field $E(t) = E_0 e^{-t^2/(2\sigma^2)}\cos(\omega_0 t)$ in resonance with the IR-mode. 
For a peak electric field of $2\,\mathrm{MV/cm}$, Eq.~\eqref{eq:absorption_bound} gives an absorbed energy of at most approximately $8\,\mathrm{meV}$ per unit cell, assuming perfect resonance, negligible damping, and no reduction of the field inside the material. In the momentum-independent approximation, this energy is shared uniformly among momenta and limits the zone-center Raman occupation to $n_R\lesssim8/196\simeq0.04$, neglecting interaction-energy corrections. By comparison, doubling the integrated Stokes intensity would require $n_R\simeq1$ if the Raman vertex remained unchanged. Restricting the coupling to fewer momenta relaxes this energy constraint, but does not increase the initial parametric gain at fixed coherent IR amplitude and zone-center coupling. Even in the favorable degenerate, undamped limit with negligible depletion of the IR oscillation, the weak-coupling estimate gives a gain scale $\Lambda=(\sqrt{5}/12)\bar\alpha^2\Delta\epsilon\lesssim0.015\,\mathrm{meV}$ for $|\bar\alpha|=0.1$. This is much smaller than the IR damping scale $\gamma_{\mathrm{IR}}\simeq0.5\,\mathrm{meV}$; the corresponding undamped evolution reaches $n_R=1$ only after approximately $30\,\mathrm{ps}$. Thus, concentrating the available energy into fewer modes does not overcome the slow transfer rate: the coherent IR drive decays before this mechanism can generate an order-unity Raman occupation.
These estimates indicate that the proposed cubic coupling, combined with an unchanged Raman vertex, cannot account for the observed intensity enhancement under the experimental conditions.\\ 
\begin{figure}[t]
     \centering
     \includegraphics[width=0.99\linewidth]{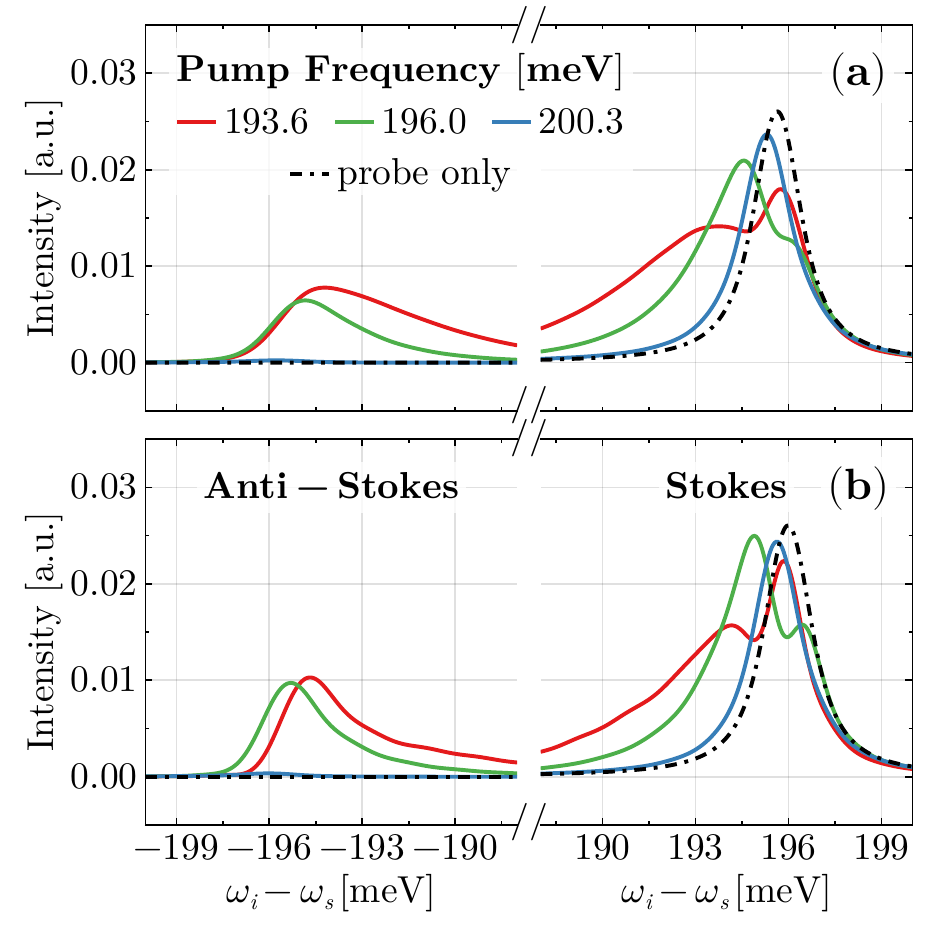}
     \caption{Nonequilibrium \eg Raman spectrum of the model Eq.~\eqref{eq:Graphite_Hamiltonian} for pump frequencies of $193.6,\, 196$ and $200.3\, \mathrm{meV}$, corresponding to red-detuned, in-resonance and blue-detuned driving, (a) using the Lindblad formalism, and (b) using the semiclassical theory.  All pump beams have a FWHM of $0.5\,\mathrm{ps}$ and a peak field strength of approximately $81\,\mathrm{MV/cm}$. The probe envelope is $4\, \mathrm{ps}$ wide and centered on the pump (zero time delay). The probe-only equilibrium Raman spectrum is shown as a dash-dotted line for comparison.}
     \label{fig:phonon-phonon_only_Raman_spectrum}
 \end{figure}
However, we can determine whether the observed optical spring effect occurs at the level of the \eg spectrum or also arises from modifications to the Raman matrix elements.
From frozen phonon calculations (Appendix \ref{sec:frozen_phonons}) we find that the model of Eq.~\eqref{eq:action} must be expanded by an additional cubic self-interaction of the \eg mode and at least one interaction of fourth order, which is the quartic self-interaction of the \eu mode. The Hamiltonian is then given by
  \begin{equation}\label{eq:Graphite_Hamiltonian}
  \begin{split}
      \hat H_\mathrm{ph} &= \omega_\R \left(\hat n_\R + \frac{1}{2} \right) + \omega_\IR \left(\hat n_\IR + \frac{1}{2} \right) \\
      &\quad + \frac{g_3^{(1)}}{2} \hat Q_\IR^2 \hat Q_\R + \frac{g_3^{(2)}}{3!} \hat Q_\R^3 + \frac{g_4^{(1)}}{4!} \hat Q_\IR^4 \,.
  \end{split}
  \end{equation}
 We neglect the remaining quartic interactions and higher-order terms because they are smaller, and among the quartic terms, $Q_\IR^4$ is by far the largest during the pump since $Q_\IR$ is driven directly. This Hamiltonian is unbounded from below and therefore, in principle, unstable. However, we do not drive the system strongly enough to probe the potential landscape outside the local minimum. For all simulations, we monitor the Hilbert-space cutoff-occupation-number overlap of the state during time evolution to control truncation error and ensure the state remains localized in the stable subspace. We choose the truncation such that this overlap never exceeds $10^{-6}$. The pump field enters as
 \begin{equation}
     \hat H(t) = \hat H_\mathrm{ph} - \abs{d^\eu_\parallel(\Omega)}  E_{d}(t) \hat Q_\IR
 \end{equation}
 with $\abs{d^\eu_\parallel(\Omega)} = 0.34 \; \mathrm{\AA}\,\sqrt{\mathrm{meV}}$, the magnitude of the mass-weighted dynamical Born effective charge of the \eu mode driven at its eigenfrequency \cite{binci_first-principles_2021}, and $E_d(t)$ the external electric field, phase-shifted by its complex phase $\arg(d^\eu_\parallel(\Omega))$, and where we chose charge units such that $e=1$.\\
  \begin{figure}[t]
     \centering
     \includegraphics[width=0.99\linewidth]{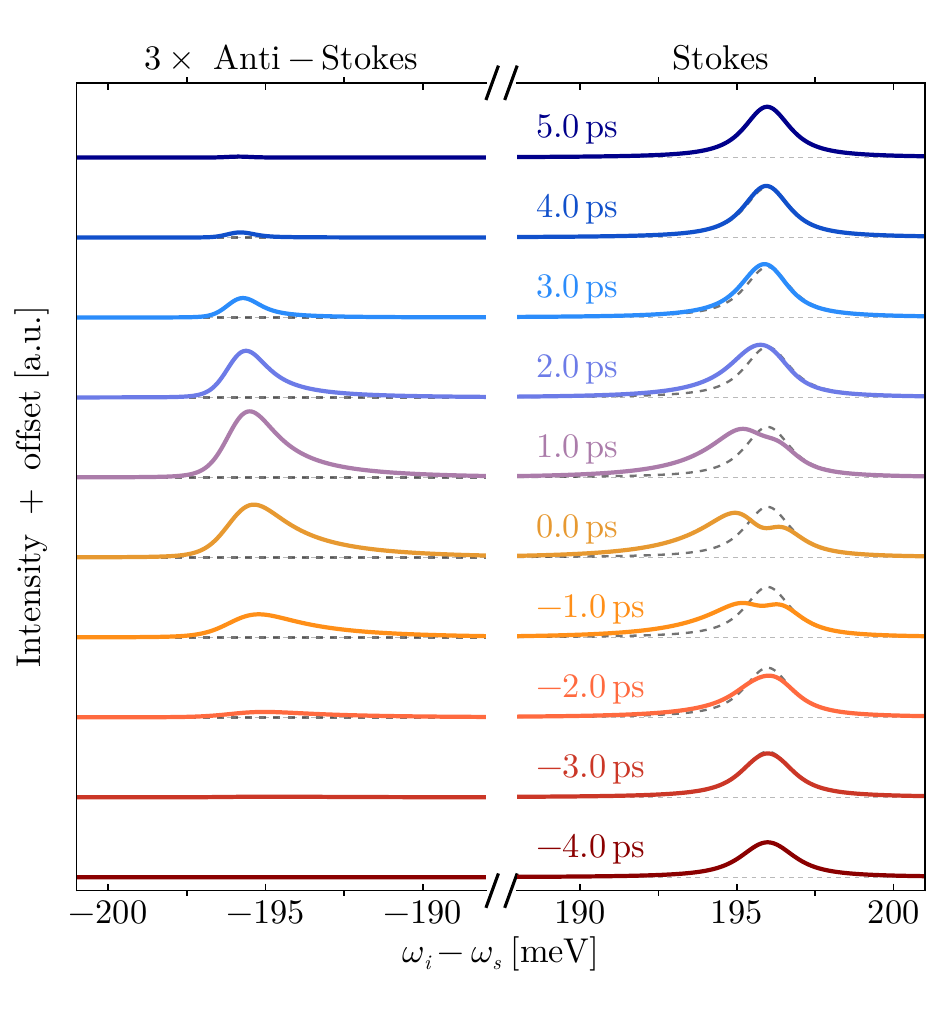}
     \caption{Pump-probe delay scan of the \eg Raman spectrum with delay time relative to a $500\,$fs FWHM pump with $81\,$MV/cm peak field-strength and $196\,$meV carrier frequency. Spectra obtained using different probe delays are offset vertically with the respective delay time labeled next to the graph using the same color. The probe width is $4\,$ps FWHM for all delays. The probe-only equilibrium \eg spectrum is indicated by a dashed line as comparison for all delay times. For improved visibility the anti-Stokes spectrum is shown threefold magnified compared to the Stokes side.}
     \label{fig:long_pulse_delay_scan}
 \end{figure}
 The nonequilibrium Raman spectra obtained from the model~\eqref{eq:Graphite_Hamiltonian} upon a $500\,$fs FWHM Gaussian pulse with $81\,$MV/cm peak field strength are shown in Fig. \ref{fig:phonon-phonon_only_Raman_spectrum} at zero pump-probe delay using both the Lindblad and semiclassical approximations. Notably, the intensity is significantly reduced compared to the simpler theory in Fig. \ref{fig:cubic_Raman_spectrum}. We found that this is predominantly due to the quartic interaction. Furthermore, we find again that for red-detuned, in-resonance, and blue-detuned pumps, the \eg mode softens. Like in the simplified model shown in Fig. \ref{fig:cubic_Raman_spectrum} the semiclassical approximation shows qualitative agreement with the full quantum treatment. 
 
\begin{figure}
     \centering
     \includegraphics[width=0.99\linewidth]{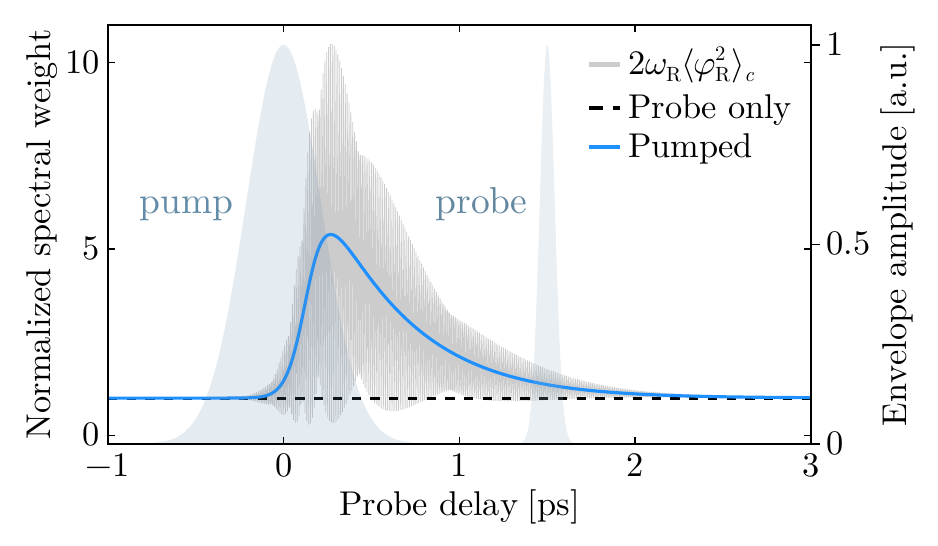}
     \caption{Integrated weight of the driven Raman spectrum (Stokes and anti-Stokes) using $100\,$fs FWHM probe pulses (envelope shown at $1.5\,$ps) against their delay compared to the pump. The pump has a carrier frequency of $\hbar\Omega = 196\,$meV, a peak field strength of $81\,$MV/cm, and a Gaussian envelope with $500\,$fs FWHM centered at $t=0$ (envelope shown at the origin). The weight is normalized to the probe-only spectrum (dashed line). We show the instantaneous \eg variance $2\omega_\R \expval{\varphi_\R^2(t)}$ of the time evolution for comparison (gray). Spectral weight is integrated between $-249\,$meV to $-144\,$meV and $144\,$meV to $249\,$meV.}
     \label{fig:short_pulse_delay_scan}
 \end{figure}

 Because the quantum theory is significantly more expensive ($\sim 10^5\times$), we present delay scans only for the semiclassical approximation. In Fig.~\ref{fig:long_pulse_delay_scan}, we show the Raman spectra calculated using~\eqref{eq:nonequilibrium_Raman_response} with $4\,$ps FWHM probe pulses centered at the indicated delay times. The spectral resolution necessary to resolve the \eg spectrum comes at the cost of time resolution. We therefore observe significant spectral modification already for probe pulses centered up to $3\,$ps before the $500\,$fs FWHM pump pulse. To achieve time resolution on the order of the pump envelope, we must sacrifice spectral resolution. Instead of plotting the smeared spectra, it is more insightful to track the integrated spectral weight. We denote the spontaneous Raman scattering intensity $I_s (\nu)$ (where $\nu = \omega_\mathrm{out} - \omega_\mathrm{in}$ is the Raman shift), the scattering intensity due to the connected correlator measured sufficiently far away from the probe reflection angle. Computationally, we obtain $I_s$ from the $t \to \infty$ limit of~\eqref{eq:nonequilibrium_Raman_response}, using only the connected correlator rather than the full correlator. Then integrating over $\nu$ one finds
 \begin{equation}
     \int_{-\infty}^\infty d\nu\, I_s(\nu) \propto \int_{-\infty}^\infty dt\, s^2(t) \expval{\varphi_\R^2(t)}_c\,.
 \end{equation}
 Hence, the spectral weight is proportional to the equal-time variance of $\varphi_\R$, weighted by the squared probe envelope, such that shorter probes yield higher time resolution. However, Rayleigh scattering limits the measurable time resolution. As the probe pulse shortens, its spectral width, as well as that of the measured Raman feature, increases. Because Rayleigh scattering is much more intense, the near edge of the frequency window over which we integrate the spectral weight should be at least 5 probe-envelope standard deviations from the Rayleigh center. Fig.~\ref{fig:short_pulse_delay_scan} shows such a time-resolved integrated spectral-weight scan using $100\,$fs probes.
Comparing Fig.~\ref{fig:long_pulse_delay_scan} with Fig.~\ref{fig:short_pulse_delay_scan}, the modest increase in spectral weight suggested by the frequency-resolved Raman measurement with a $4\,$ps FWHM probe is due to the short pump and, consequently, the short-lived excited state relative to the probed time scale. The time-resolved spectral-weight evolution shows the cycle-averaged amplitude variance increasing by more than fivefold. In turn this suggests that to generate the observed transient signal with a $4\,$ps probe, the incoherent occupation of the Raman mode shortly after the pump would have needed to exceed unity substantially, further corroborating our previous estimates.
 
It is therefore probable that both the intensity gain and the resonance shift in the experiment are caused predominantly by pump-induced modification of the Raman tensor, that is, on the level of electron-phonon coupling rather than by genuine nonlinear phononics. Although its magnitude remains to be established, enhancing the Raman matrix elements by a factor of $\sqrt{2}$, i.e., a $41\%$ increase, would reproduce the doubling of the Stokes intensity without any gain in the \eg occupation. The near-resonance dependence of the pump on the \eu mode suggests that the coherent \eu oscillation modifies the Raman vertex. Based on crystal symmetry, we find the expansion
\begin{equation}
\begin{split}
    R_{ij,\nu} &\approx R_{ij,\nu}^{(0)} + C^{(QQ)}_{ij,\nu;kl}\,Q_{\IR,k} Q_{\IR,l}\\
    &\quad+ C^{(QE)}_{ij,\nu;kl}\,Q_{\IR,k} E_l + C^{(EE)}_{ij,\nu;kl}\,E_k E_l + \mathcal{O}_4
\end{split}
\end{equation}
where $\mathcal{O}_4$ denotes terms of total degree four or higher. Previous work observed a threefold increase in the electron-phonon coupling parameter upon resonant driving of the \eu mode with a fluence of only $0.3\,$mJ/cm$^2$ in bilayer graphene \cite{pomarico_enhanced_2017}. In particular, the corrections to $R_{ij}^{(0)}$ inherit time-dependence from the oscillating phonon and field amplitudes. Thus, we must include $R_{ij}$ in the integral in~\eqref{eq:nonequilibrium_Raman_response}, which introduces additional pump-frequency dependence. However, determining the expansion coefficients is beyond the scope of this article. 
We therefore distinguish changes in the underlying phonon correlation function from changes in the Raman matrix elements when interpreting the transient spectral response.

\section{Conclusion}
We have developed a framework for quantum nonlinear phononics that connects the dynamics of coherently driven, interacting phonons to their nonequilibrium Raman response. Although the relative quantum uncertainty of a macroscopically driven mode vanishes in the thermodynamic limit, its coherent amplitude does not fully characterize the driven lattice. Anharmonic interactions endow phonon fluctuations with independent dynamics, including squeezing, and allow them to feed back on the mean displacement. These contributions remain finite as the system size increases and cannot, in general, be absorbed into a static renormalization of the classical equations of motion. A self-consistent Gaussian approximation captures the leading interplay between coherent motion and fluctuations, while comparison with quantum master-equation calculations establishes its qualitative usefulness and quantitative limitations.

Time-resolved spontaneous Raman scattering provides direct experimental access to driven fluctuations. We have expressed the nonequilibrium Raman intensity in terms of the two-time phonon correlator and distinguished its connected fluctuation contribution from the disconnected contribution of coherent motion. Their different angular emission patterns provide a way to isolate the fluctuation spectrum from coherent Raman sidebands. This distinction is essential for interpreting nonequilibrium Raman spectra: changes in spontaneous Raman intensity and peak position probe the fluctuation spectrum, and an enhanced signal need not imply thermal heating. On the other hand, information on the mean-field oscillation can be obtained from the coherent sidebands.

For the nearly degenerate \eu and \eg optical phonons of AB-stacked multilayer graphene, we find that driving the IR-active \eu mode modifies the quantum fluctuation and Raman spectrum of the \eg mode substantially for large pump fluences. However, anharmonic couplings informed by frozen-phonon calculations alone yield softening for both red- and blue-detuned pumping and therefore do not reproduce the experimentally observed reversal of the frequency shift. Furthermore, the discrepancy in pump intensity between our theory and experiment, where significant transient Raman intensity is observed, suggests that the Raman tensor and electron-phonon couplings more generally play an important role in the material's pump-induced dynamics.

More broadly, these findings extend the description of nonlinear phononics beyond control of average atomic positions to the dynamics of lattice fluctuations. In experiments seeking to induce electronic, magnetic, or structural phases through selective phonon excitation, coherent displacements and fluctuation dynamics should therefore be assessed together rather than inferred from a classical trajectory alone. Extending the present treatment to dispersive phonons is feasible using the semiclassical approximation we presented, and their coupling to electronic degrees of freedom will help determine how these distinct contributions influence transient order. Pump-probe Raman scattering offers a route to test this expanded picture and establish the role of quantum lattice dynamics in light-induced states of matter.\\

\section*{Acknowledgement}
We thank Andrea Cavalleri, Paulo Forni, James Freericks, Tomke Glier, Henrik Müller-Groeling, Sida Tian, Mohsen Yarmohammadi, and Carl Zelle for helpful discussions. We thank the
Max Planck-UBC-UTokyo Center for Quantum Materials for fruitful collaborations and financial support. AI use disclosure: Gemini-3.1 Pro was used to generate, optimize, and debug large parts of the code. We verified the generated code before use. ChatGPT-6 Astra was used for internal review before submission.

\bibliography{Phononpaper}

\appendix
\crefalias{section}{appendix}

 \section{Feynman diagram notation}\label{app: Feynman_rules}
On all three levels of dressing, we use zigzags to denote the Raman phonon and plain lines for the IR phonon.

\begin{fmffile}{Definitions}
    \begin{align*}
    G_{0,\IR} &= \;\eqgraph{0.5 ex}{0ex}{\begin{fmfgraph*}(10,3)
        \fmfleft{i}
        \fmfright{o}
        \fmf{plain,width=1}{i,o}
    \end{fmfgraph*}}\\[-4.5ex]
    & & 
    -\alpha &= \eqgraph{1ex}{0ex}{
    \begin{fmfgraph}(10,10)
        \fmfleft{i1,i2}
        \fmfright{o}
        \fmf{plain}{i1,v1}
        \fmf{plain}{i2,v1}
        \fmf{zigzag,tension=1.5}{v1,o}
        \fmfdot{v1}
    \end{fmfgraph}}\\[-4.5ex]
    G_{0,\R} &= \;\eqgraph{1ex}{0ex}{\begin{fmfgraph*}(10,3)
        \fmfleft{i}
        \fmfright{o}
        \fmf{zigzag,width=1}{i,o}
    \end{fmfgraph*}}\\
\left(\Gamma^{(2)}_{\IR}\right)^{-1} &= \;\eqgraph{1ex}{0ex}{\begin{fmfgraph*}(10,3)
        \fmfleft{i}
        \fmfright{o}
        \fmf{plain,width=2}{i,o}
    \end{fmfgraph*}}&
    \qquad G_{\IR} &= \;\eqgraph{1ex}{0ex}{\begin{fmfgraph*}(10,3)
        \fmfleft{i}
        \fmfright{o}
        \fmf{dbl_plain,width=1}{i,o}
    \end{fmfgraph*}}\\
    \left(\Gamma^{(2)}_{\R}\right)^{-1} &= \;\eqgraph{1ex}{0ex}{\begin{fmfgraph*}(10,3)
        \fmfleft{i}
        \fmfright{o}
        \fmf{zigzag,width=2}{i,o}
    \end{fmfgraph*}}&\qquad
    G_{\R} &= \;\eqgraph{1ex}{0ex}{\begin{fmfgraph*}(10,3)
        \fmfleft{i}
        \fmfright{o}
        \fmf{dbl_zigzag,width=1}{i,o}
    \end{fmfgraph*}}\\
    \Gamma^{(2)}_{\IR\R} &= \;\eqgraph{1ex}{0ex}{\begin{fmfgraph*}(10,3)
        \fmfleft{i}
        \fmfright{o}
        \fmf{plain,width=1}{i,v1}
        \fmf{zigzag,width=1}{v1,o}
        \fmfv{d.sh=circle,d.f=empty,d.si=10,l.d=0}{v1}
    \end{fmfgraph*}}&\qquad
    G_{\IR\R} &= \;\eqgraph{1ex}{0ex}{\begin{fmfgraph*}(10,3)
        \fmfleft{i}
        \fmfright{o}
        \fmf{dbl_plain,width=1}{i,v1}
        \fmf{dbl_zigzag,width=1}{v1,o}
    \end{fmfgraph*}}
\end{align*}
\end{fmffile}

\section{Linear level-repulsion}
An even simpler model for the observation of \cite{buchenau_optically_2023}, relying entirely on classical theory, would be to consider only the linear light-mediated coupling $c_1 F(t) X_\R X_\IR$. A coherent light beam breaks the inversion symmetry of the crystal by distinguishing a direction via its electric field. This allows the even and odd modes to interact linearly. In this case, the Raman- and IR-active modes hybridize into mixed modes
\begin{equation}
    \begin{split}
        X_+ &= -\sin(\theta) X_\R + \cos(\theta) X_\IR\\
        X_- &= \cos(\theta) X_\R + \sin(\theta) X_\IR
    \end{split}
\end{equation}
with repelled eigenenergies, 
\begin{equation}
    \omega^2_\pm = \frac{\omega_\R^2 + \omega_\IR^2}{2} \pm \frac{1}{2}\sqrt{\Delta^2 + 4c_1^2F(t)^2}\,.
\end{equation}
Here $\tan(2\theta) = - \frac{2c_1 F(t)}{\Delta}$, and $\Delta := \omega_\IR^2 - \omega_\R^2$ is the mode detuning. The weights explicitly evaluate to
\begin{align}
\begin{split}
    \cos\theta &= \sqrt{\frac{1}{2}\left(1+\frac{\Delta}{\sqrt{\Delta^2 + 4c_1^2F(t)^2}}\right)}\quad\text{and}\\
    \sin\theta &= -\text{sgn}(c_1 F(t))\sqrt{\frac{1}{2}\left(1-\frac{\Delta}{\sqrt{\Delta^2 + 4c_1^2F(t)^2}}\right)}\,.
\end{split}
\end{align}
 Both hybridized modes are Raman active owing to their finite \eg character and should appear as two distinct peaks in the Raman spectrum. The Raman intensities of the two hybridized modes are proportional to the squares of their respective $X_\R$ characters, that is, $\sin^2\theta$ and $\cos^2\theta$ for $X_+$ and $X_-$, respectively. The finite detuning $\Delta$ allows for an asymmetric hybridization such that the appearance of only a single peak in the experiment could be explained by one hybridized mode being overwhelmingly of $X_\R$ character and the other being mostly of $X_\IR$ character. However, the observed peak shift during the IR-pump of $1\,\mathrm{meV}$ would imply a ratio of only approximately $2:1$ between the respective Raman intensities. Furthermore, in this scenario, the Raman peak should always soften regardless of the driving frequency, whereas in the experiment, softening is observed for a red-detuned pump and hardening for a blue-detuned pump \cite{buchenau_optically_2023}. Consequently, this mechanism is insufficient to explain the finding.\\

\section{Frozen Phonon Computation}\label{sec:frozen_phonons}
\begin{figure}
     \centering
     \includegraphics[width=0.99\linewidth]{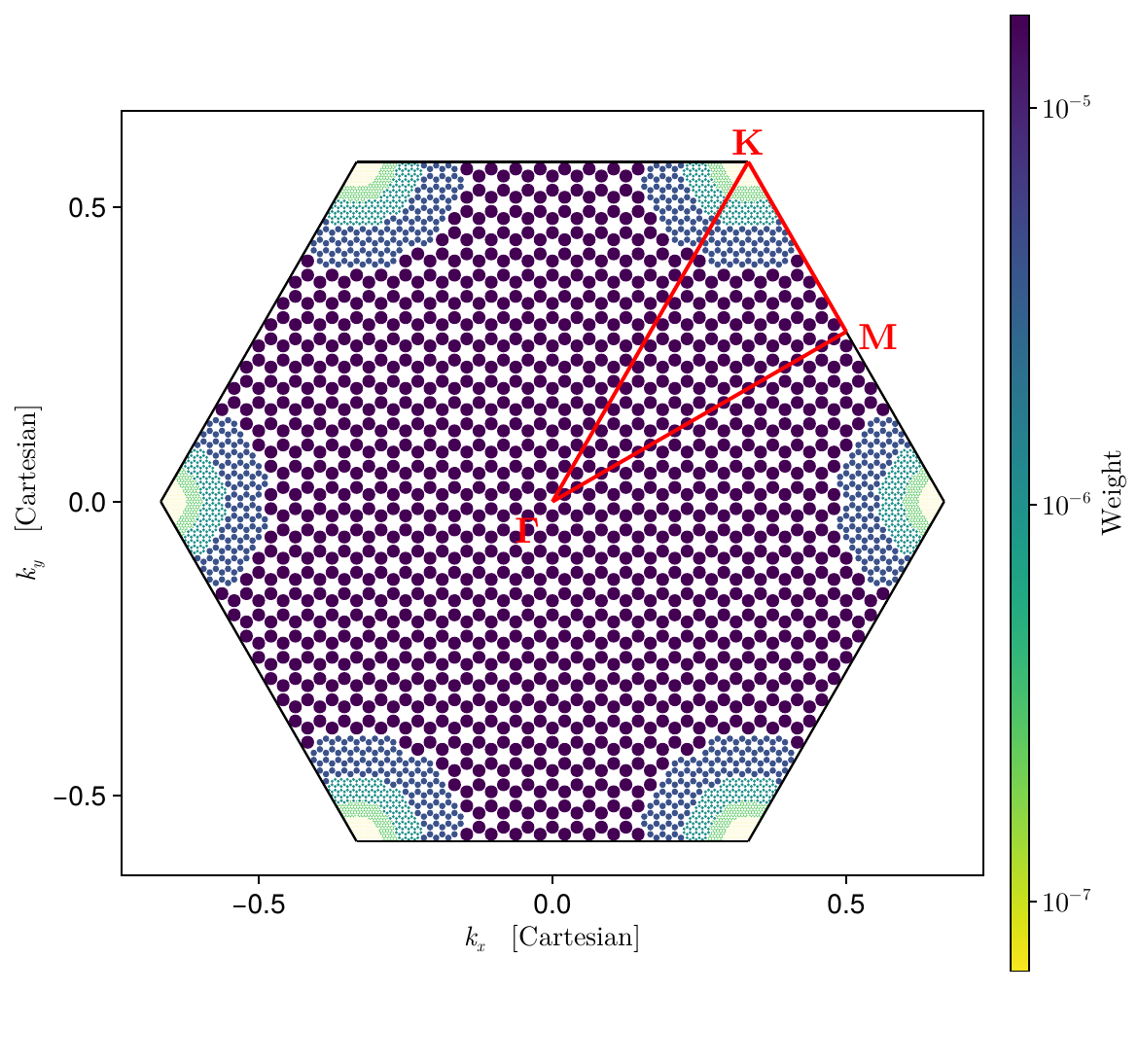}
     \caption{Nonuniform grid used for the potential energy surface calculation. We use very dense sampling on the Fermi surface around the $K-H$ line. The sampling density at level $l$ is $4^l/A_0$, where $A_0$ is the TRS-BZ area. By construction, the $l=9$ level covers a disk of radius $R_\textrm{core} = 0.1 \textrm{ \AA}^{-1}$ to capture the equilibrium Fermi surface, its shift due to a small phonon displacement, and the resonant manifold around it.}
     \label{fig:k_mesh}
 \end{figure}

To derive the interaction potential between the \eg and \eu phonons, we used the frozen-phonons method introduced in \cite{parlinski_first-principles_1997} and first applied to nonlinear phononics in \cite{subedi_theory_2014}. It entails deriving the phonon-phonon interaction by calculating the ground-state energy of the material with its atomic positions displaced to varying degrees from equilibrium according to the phonons' normal-mode vectors. For the amplitudes of the \eg and \eu modes' two polarizations, that is, for the four coordinates
\begin{equation}
    X_\eu = \begin{pmatrix}
    u_x \\ u_y
\end{pmatrix}\quad \text{and} \quad X_\eg = \begin{pmatrix}
    g_x \\ g_y
\end{pmatrix}\,.
\end{equation}
 we discretized the interval $-0.02\,$\r{A} to $0.02\,$\r{A} into 5 points; hence, the total grid measures $625$ points. At each point, we compute the ground-state energy via DFT using the open-source software Quantum ESPRESSO \cite{giannozzi_advanced_2017}. Extracting the couplings from the total energy is difficult because small atomic displacements change the total ground-state energy only slightly, so the self-consistent field calculation requires very strict convergence thresholds. Unlike insulators, where small atomic displacements cannot close the electronic gap and therefore cannot change which electronic states are occupied, metals and semimetals like graphite can have their Fermi surface reshaped. As a result, we need to resolve the momentum-space area around the Fermi surface in detail. To resolve the needle-shaped Fermi surface of graphite around the $K-H$ line, we adopted the custom momentum grid approach described in \cite{binci_first-principles_2021} (see figure \ref{fig:k_mesh}), where we sample a relatively large cylinder around the $K-H$ line of radius $0.1\,\mathrm{\AA}^{-1}$ ($\mathcal{N}=17$) with $l=9$ triangulation recursion to account for the shifting of the Dirac point due to the small phonon displacement. Locally, this matches the resolution of a uniform $724\times 724\times 40$ grid (we use $N_{k_z} = 40$). The density subsequently decays cubically to a minimum triangulation level of $l_\mathrm{min}=4$. Because the frozen-phonon calculation displaces the atomic positions by phonon eigenvectors, the full D$_{6h}$ symmetry is broken. Therefore, we cannot use the irreducible Brillouin zone (IBZ) (the wedge in figure \ref{fig:k_mesh}); instead, we must sample the entire time-reversal-symmetrized Brillouin zone (TRS-BZ), which is six times larger. We performed the DFT calculation with Fermi-Dirac smearing and 'degauss = 0.002', corresponding to a temperature of $\approx 27$ meV and an energy convergence threshold of $10^{-10}\,$ Ry. We estimate the statistical error of the quartic couplings derived from the DFT error via error propagation as $0.06\, \mathrm{eV}/\mathrm{\AA}^4$, which is negligible compared to the systematic errors.
  We used the resulting energy surface to fit the interaction potential via least squares. The fitting polynomial includes all symmetry-allowed interactions up to fourth/sixth order. These can be systematically derived in the chiral phonon basis
\begin{equation}\label{eq:chiral_basis}
    G := g_x + ig_y\qquad U := u_x + i u_y
\end{equation}
by collecting all terms that are invariant under all D$_{6h}$ symmetry operations, generated by the three operations, inversion ($i$), $\pi/3-$rotation ($C_6$), and vertical reflection ($\sigma_v$) under which the coordinate vectors transform as follows (we choose the basis where a carbon-carbon bond is aligned with the y-axis):
\begin{align}
    i:\quad& G \mapsto G,\quad U\mapsto -U
    \\
    C_6:\quad& G \mapsto e^{-i2\pi/3} G,\quad U\mapsto e^{i\pi/3} U
    \\
    \sigma_v:\quad& G \mapsto -\bar G,\quad U\mapsto -\bar U
\end{align}
where complex conjugation is denoted with a bar. Then up to the fourth order one finds
    \begin{equation}
    \begin{split}
        V(U,G) &\approx \frac{k_u}{2}\abs{U}^2 + \frac{k_g}{2}\abs{G}^2  + \frac{g_3^{(1)}}{2} \Im (U^2 G)\\
        &\quad+ \frac{g_3^{(2)} }{3!}\Im (G^3) + \frac{g_4^{(1)}}{4!} \abs{U}^4 + \frac{g_4^{(2)}}{4!} \abs{G}^4\\
        &\quad  + \frac{g_4^{(3)}}{4} \abs{U}^2 \abs{G}^2 + \frac{g_4^{(4)}}{4} \Re(U^2\bar G^2)\,.
    \end{split}
\end{equation}
The result of the fit is presented in Table \ref{tab:DFT_couplings}. To estimate the systematic error introduced by truncating the potential at the fourth order, we also perform a fit up to sixth order. The numbers displayed in brackets in Table \ref{tab:DFT_couplings} are the difference to this sixth-order fit and do not include the systematic errors of the frozen phonon method and the DFT approximation.\\
Notably, using a uniform $30\times 30\times 20$ Monkhorst-Pack grid (not resolving the Fermi surface) we found significantly different values for the quartic couplings, that is $g_4^{(1)}  \approx 250\, \mathrm{eV/\AA}^4$, $g_4^{(2)}  \approx 260\, \mathrm{eV/\AA}^4$, $g_4^{(3)} \approx  2400\, \mathrm{eV/\AA}^4$ and $g_4^{(4)} \approx  2100\, \mathrm{eV/\AA}^4$ corresponding to $\bar g_4^{(1)}  \approx 0.004 \approx \bar g_4^{(2)}$, $\bar g_4^{(3)}  \approx 0.041$ and $\bar g_4^{(4)}  \approx 0.034$. The cubic parameters differ by only a few percent from those obtained using the non-uniform grid. Convexity of the potential is provided by positive sixth-order coefficients.
\begin{table}[htbp]
    \caption{Couplings derived from the frozen phonon method}
    \label{tab:DFT_couplings}
    \renewcommand{\arraystretch}{1.5}
    \begin{tabular*}{\linewidth}{@{} l @{\extracolsep{\fill}} c r @{}}
        \toprule
        Parameter & Nominal Value & Dimensionless Value\\
        \midrule
        $k_u$ & $108.19 \,(+ 0.03) $ eV/\AA$^2$ & $1.0000$ \\
        $k_g$ & $107.30 \,(- 0.01) $ eV/\AA$^2$ & $0.9918$ \\
        $g_3^{(1)}$ & $-268.5 \,(- 1.3)$ eV/\AA$^3$ & $-0.1050$ \\
        $g_3^{(2)}$ & $-258.0 \,(+ 1.2)$ eV/\AA$^3$ & $-0.1009$ \\
        $g_4^{(1)}$ & $-1900 \,(- 1200)$ eV/\AA$^4$ & $-0.0319$ \\
        $g_4^{(2)}$ & $41.5 \,(+ 0.8)$ eV/\AA$^4$ & $0.0007$ \\
        $g_4^{(3)}$ & $142 \,(+ 26)$ eV/\AA$^4$ & $0.0024$ \\
        $g_4^{(4)}$ & $55 \,(+ 4)$ eV/\AA$^4$ & $0.0009$ \\
        \bottomrule
    \end{tabular*}
\end{table}
The dimensionless values are obtained by first transforming the nth-order couplings in units of eV/\AA$^n$ to couplings between our mass-weighted normal coordinates in units of meV via multiplication with $(0.589)^n 10^3$. Here we used that both mode masses are given by the mass of carbon and using length units of \AA,  energy units of meV and $\hbar = 1$, we get the conversion factor $\hbar/\sqrt{m_C} \approx 0.589\, \sqrt{\mathrm{meV}}\mathrm{\AA}$ between mass-weighted and not mass-weighted normal mode coordinates. The value we thus obtain for $k_u$ defines $\omega_\IR^2$. The dimensionless values are then obtained from dividing by the appropriate power of $\omega_\IR$ as the reference scale.

\section{Field-mediated interactions}\label{sec:field_mediated_couplings}
In addition to interactions purely among the phonon branches, field-mediated phonon-phonon interactions (photophononic interactions) can be relevant in the driven state~\cite{xu_time-domain_2025}. For a normal-incident pump beam, the electric field transforms in the \eu irrep. Therefore, we can construct light-phonon-phonon interactions by replacing any number of \eu amplitudes in pure phonon-phonon interactions with an electric field.
Given electric field strengths weaker than the interatomic forces, the coupling strength decreases rapidly with the exponent of the electric field. Hence, we consider interactions that are linear in the electric field. In particular, we add the following light-mediated phonon-phonon interactions to the Hamiltonian:
 \begin{equation}\label{eq:Vlmc_simplified}
     \hat V_\mathrm{lmc}(t) = \abs{g_1} \hat Q_\R \hat Q_\IR E_{g_1}(t) + \frac{\abs{g_2}}{2} \hat Q_\R^2 \hat Q_\IR E_{g_2}(t)
 \end{equation}
 with $E_{c_{1/2}}(t)$ denoting the electric field phase-shifted by the complex phase of the respective coupling $\arg(c_{1/2})$.
 The couplings depend on two/three frequencies for the cubic/quartic terms. However, we are interested only in their values at the driving frequency of the electric field and in the $\omega \to 0$ limit with respect to the Raman mode. Hence, they are treated as constants $g_1 = g_1 (\Omega_\mathrm{drive},0)$, $g_2 = g_2 (\Omega_\mathrm{drive},0,0)$.\\ 
 Light-mediated interactions lead to distinct phenomenology owing to their sensitivity to phase shifts between the driving field and the phonons. To illustrate this, suppose $F(t) \sim \sin(\Omega t)$ drives the \eu mode $Q_\IR(t) \sim \bar Q_\IR \sin(\Omega t - \theta)$ where the phase depends on the damping parameter of the \eu mode and shows the characteristic phase shift of a driven damped harmonic oscillator $\theta = \mathrm{atan2}[\gamma \Omega,(\omega_0^2 - \Omega^2)]$.
 The effect of this relative phase is qualitatively distinct between the purely phononic interaction $Q_\IR^2Q_\R$, where $Q_\IR(t)^2 \sim 1 - \cos(2\Omega t - 2\theta)$, and the mixed interactions $Q_\IR Q_\R^{(2)} F$, where $Q_\IR(t)F(t) \sim \cos(\theta) - \cos(2\Omega t - \theta)$.
The dependence of the difference-frequency-generation (DFG) on the relative phase between the \eu mode and the pump laser enables the optical spring effect observed in the experiment. Both directly, via the quartic interaction $\abs{g_2} \hat Q_\R^2 \hat Q_\IR E_{g_2}(t)$ producing a phase-dependent modification of the \eg pole:
\begin{equation}
    \omega_\R^2 \rightarrow \omega_\R^2 + \frac{\abs{g_2}E_0 \bar Q_\IR}{2} \cos(\theta - \theta_{g_2})
\end{equation}
and indirectly via the cubic interaction $\abs{g_1} \hat Q_\R \hat Q_\IR E_{g_1}(t)$ generating a phase-dependent DC force on the \eg mode, which is translated to a modification of the resonance frequency via the phonon inherent anharmonicity $\sim g_3^{(2)}\hat Q_\R^3$. A priori, it is unclear which is more relevant and whether either effect is strong enough to explain the experimental finding.\\ 
The simplified field-mediated interactions~\eqref{eq:Vlmc_simplified} take the form
\begin{equation}\label{eq:V_lmc}
\begin{split}
    V_\mathrm{lmc} &= c_1 \Im\left(EUG\right)\\
    &\quad + c_{2}^{(a)} \Re\left(E \bar U\right)\abs{G}^2 + c_{2}^{(b)} \Re\left(EU \bar G^2\right)\,,
\end{split}
\end{equation} 
when polarizations are taken into account, using the chiral basis~\eqref{eq:chiral_basis}.\\
A phonon mode $\nu$ couples to the electric field via its dynamical Born effective charge (BEC) $Z^*_{\alpha,\nu}(\omega)$, encoding the polarization in the direction $\alpha$ induced by the displacement of the phonon $\nu$ \cite{binci_first-principles_2021,dreyer_nonadiabatic_2022,marchese_born_2024}. In metals, this property is frequency-dependent because the valence electrons near the Fermi surface respond non-adiabatically to the atomic displacement. In an effective formulation where the valence electrons have been integrated out, this leads to a frequency-dependent interaction between the phonons and the external field, which is expressed most naturally in the action formalism:
\begin{equation}\label{eq:BEC_action}
    S_{ph-E} = \int_\omega eE_\alpha(\omega)Z^*_{\alpha,\nu}(\omega)X_\nu(-\omega)\,.
\end{equation} 
In the Cartesian basis, the BEC tensor is defined as the change of the polarization direction $\alpha$ at time $t$ with respect to the displacement of atom $\kappa$ in direction $\beta$ at time $t'$,
\begin{equation}
    Z^*_{\alpha,\kappa\beta}(t,t') = \frac{\Omega}{e} \frac{\partial P_\alpha(t)}{\partial X_{\kappa \beta}(t')}\,.
\end{equation}
 Causality implies that $\Re Z(-\omega)=\Re Z(\omega)$ and $\Im Z(-\omega)=-\Im Z(\omega)$.
The mode-specific BEC is then obtained by contraction with the mode's eigenvector $e^\nu_{\kappa\beta}$ for mode $\nu$:
\begin{equation}
    Z^*_{\alpha,\nu} = \sum_{\kappa\beta} Z^*_{\alpha,\kappa\beta} e^\nu_{\kappa\beta}\,.
\end{equation}
When using mass-weighted normal coordinates $Q_\nu = \sqrt{M_\nu}\, X_\nu$, where $M_\nu = (\sum_{\kappa\beta}\vert e^\nu_{\kappa\beta}\vert^2/m_\kappa)\inv$ is the mode-effective mass, it is natural to define the \emph{oscillator strength} \cite{binci_first-principles_2021}
\begin{equation}
    d_{\alpha,\nu} = \sum_{\kappa} Z^*_{\alpha,\kappa\beta}\frac{e^\nu_{\kappa\beta}}{\sqrt{m_\kappa}}
\end{equation}
which takes the place of $Z^*_{\alpha,\nu}$ in~\eqref{eq:BEC_action} when $Q_\nu$ is used instead of $X_\nu$.
The coupling parameters $c_1$, $c_2^{(a)}$ and $c_2^{(b)}$ of the light-mediated interactions are obtained as the expansion coefficients of the \eu mode dynamical BEC, $Z^U_{\alpha,i}$ where $i$ indexes the two \eu polarizations, in the \eg mode displacement
\begin{equation}
    Z^U_{\alpha,i}(g_x,g_y) = Z^{U,0}_{\alpha,i} + \frac{\partial Z^U_{\alpha,i}}{\partial g_j} g_j + \frac{1}{2}\frac{\partial^2 Z^U_{\alpha,i}}{\partial g_j \partial g_k} g_j g_k\,,
\end{equation}
where we suppressed the frequency arguments and the derivatives are evaluated at $g_x = 0 = g_y$.
By expanding the symmetry channels in~\eqref{eq:V_lmc} we arrive at
\begin{widetext}
\begin{equation}\label{eq:displacement_dependent_BEC_matrix}
    Z^U(g_x,g_y) = \begin{pmatrix}
        Z^0_{xx} - c_1 g_y - (c_{2}^{(a)} + c_{2}^{(b)}) g_x^2 - (c_{2}^{(a)} - c_{2}^{(b)}) g_y^2 & Z^0_{xy} - c_1 g_x - 2 c_{2}^{(b)} g_x g_y \\
        Z^0_{yx} - c_1 g_x - 2 c_{2}^{(b)} g_x g_y & Z^0_{yy} + c_1 g_y - (c_{2}^{(a)} - c_{2}^{(b)}) g_x^2 - (c_{2}^{(a)} + c_{2}^{(b)}) g_y^2
    \end{pmatrix}\,.
\end{equation}
\end{widetext}
In writing $Z^U$ as a $2\times 2$ matrix, we only consider the in-plane electric field polarizations $\alpha = x,y$  because the z-components vanish.
We can extract the couplings by displacing only $g_x$. In general, the couplings depend on two/three frequencies for the cubic/quartic terms. However, because we are primarily interested in the DFG channel, we can use the frozen-phonon method to displace the \eg mode. We can therefore obtain the relevant couplings by fitting the dependence of the \eu mode's dynamical BEC on a static displacement of the \eg mode amplitude.
To this end, the dynamical Born effective charge is obtained via a simplified version of Eq. (7) in \cite{binci_first-principles_2021} derived via time-dependent perturbation theory, where we use the independent particle approximation. 
\begin{equation}\label{eq:BEC_TDDFPT}
    Z^{el}_{\alpha,\kappa\beta}(\omega) = -\frac{2i}{N_k} \sum_{n,m,k} \frac{\left(f_{nk}-f_{mk}\right) v_{mn}^\alpha(k) g^{\kappa\beta}_{nm}(k,0)}{(E_{nk}- E_{mk})^2 - (\hbar \omega + i\eta)^2}
\end{equation}
with $N_k$ the number of k-points, $f_{nk}$ the Fermi-Dirac function at $E_{nk}$, $v_{mn}^\alpha(k) = \bra{m,k}\hat v^\alpha \ket{n,k}$ the velocity matrix element and $g^{\kappa\beta}_{nm}(k,q) = \bra{n,k+q}\partial_{\tau_{\kappa\beta}(q)} V_{SC}\ket{m,k}$ the electron-phonon vertex matrix element in the cartesian basis where $\kappa$ runs over the carbon atoms in the unit cell and $\beta$ denotes the cartesian displacement direction. Following \cite{binci_first-principles_2021}, we estimated the electronic damping parameter $\eta$ to be one order of magnitude smaller than the evaluation frequency; hence, we set $\eta = 20\,\mathrm{meV}$. The full dynBEC is obtained by combining the electronic contribution $Z^\mathrm{el}
_{\alpha,\kappa\beta}$ with the nuclear (pseudo)charge $Z_\kappa$ of atom $\kappa$ ($Z_\kappa = 4$ in the case of carbon), that is $Z^I_{\alpha,\kappa\beta} = Z_\kappa \delta_{\alpha\beta}$ and in total \cite{wang_dynamical_2022,binci_first-principles_2021}
\begin{equation}
    Z^*_{\alpha,\kappa\beta} = Z^\mathrm{el}
_{\alpha,\kappa\beta} + Z^I_{\alpha,\kappa\beta}
\end{equation}\\
\begin{figure}
    \centering
    \includegraphics[width=0.99\linewidth]{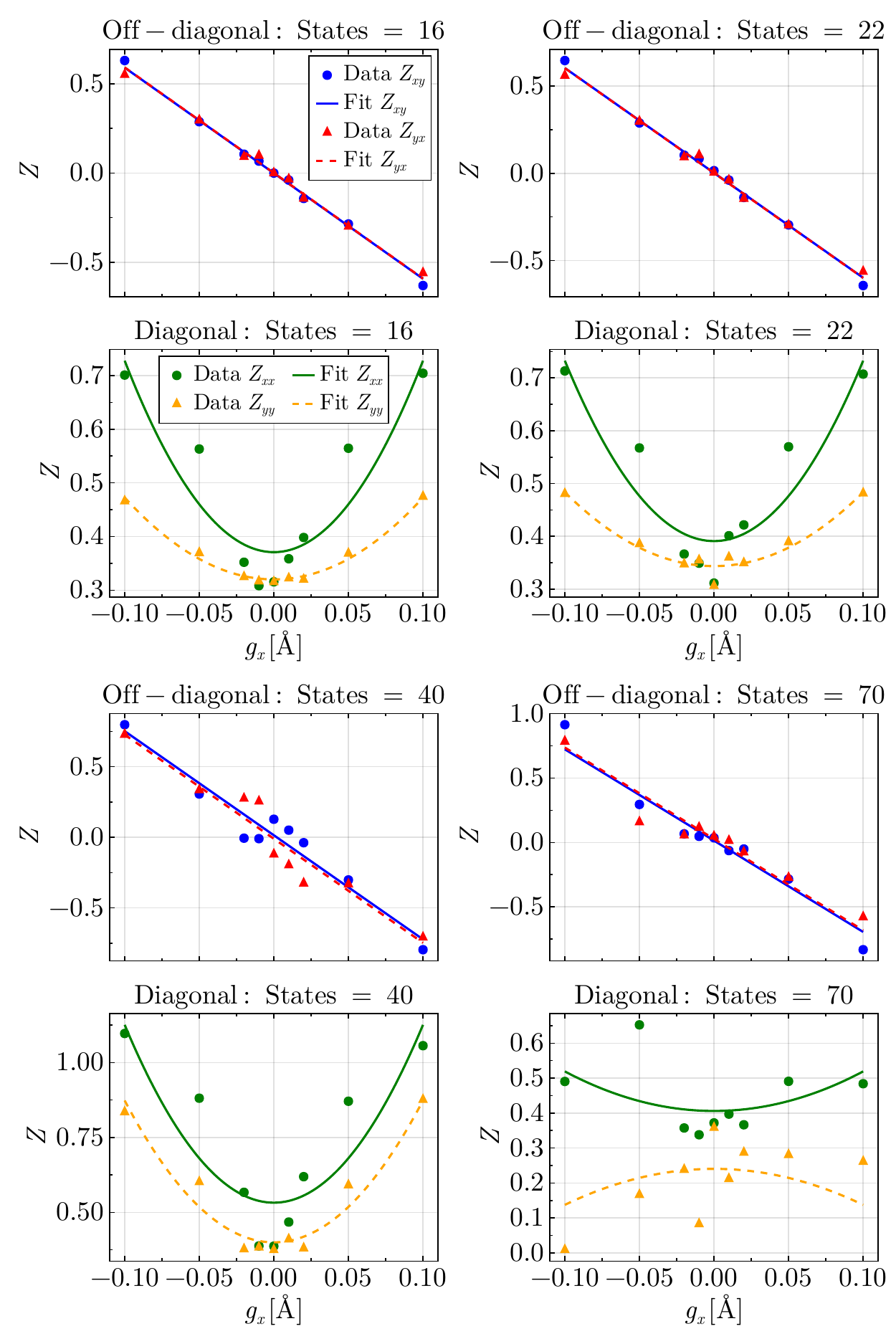}
    \caption{Fits of the \eu mode's dynamic Born effective charge variation with the \eg mode $x-$polarization displacement according to Eq.~\eqref{eq:displacement_dependent_BEC_matrix}. The fits are shown for the data obtained by using 16, 22, 40, and 70 bands. In each of the four cases, the fit of the off-diagonal matrix elements, which determine $c_1$, is shown in the upper panel and the fit for the diagonal elements, determining $c_2^{(a)}$ and $c_2^{(b)}$, is shown in the lower panel.}
    \label{fig:lmc_displacement_fit}
\end{figure}
\begin{figure}
    \centering
    \includegraphics[width=0.99\linewidth]{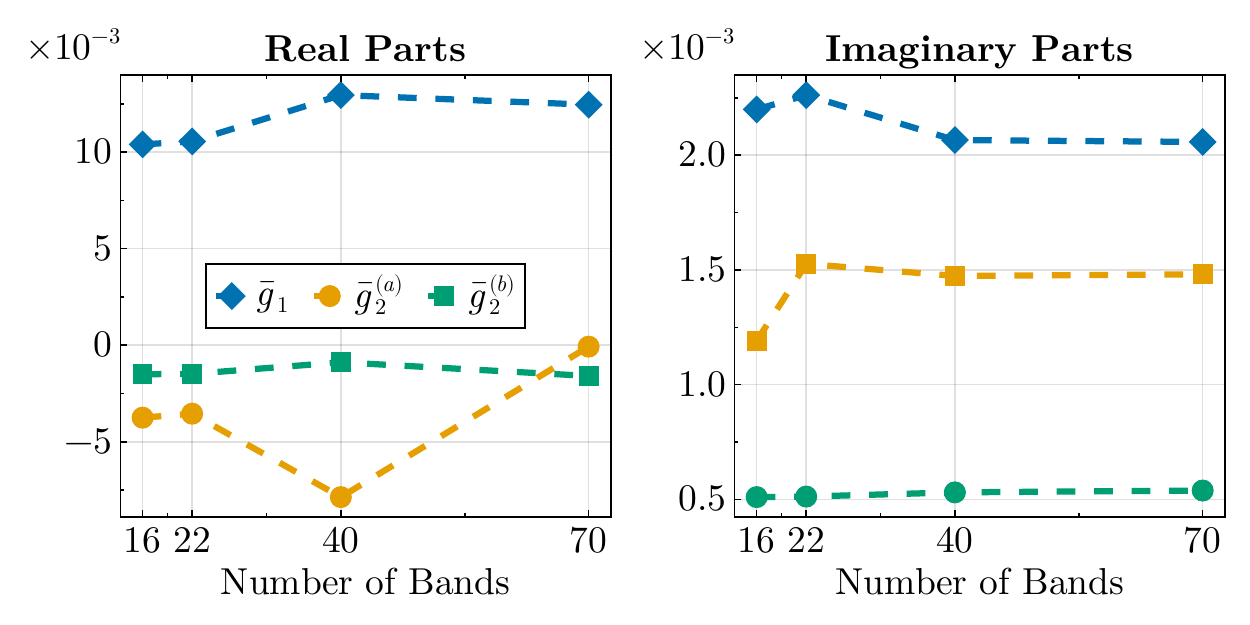}
    \caption{Real parts (left) and imaginary parts (right) of the three light-mediated coupling coefficients obtained by fitting the dynamical Born effective charge variation with \eg phonon displacement (see Fig.~\ref{fig:lmc_displacement_fit}) vs. the number of bands used in the sum over states formula~\eqref{eq:BEC_TDDFPT}. To make the coupling strengths comparable, we show their "dimensionless" values, i.e., in units of $\mathrm{\AA}\omega_\IR$ and $\mathrm{\AA}\omega_\IR^{3/2}$ for $\bar g_1$ and $\bar g_2$ respectively.}
    \label{fig:lmc_coupling_convergence}
\end{figure}
We obtain the electron-phonon and velocity matrix elements via DFPT using the EPW software package, computing the matrices in the basis of maximally localized Wannier functions (MLWFs) \cite{giustino_electron-phonon_2017, lee_electronphonon_2023}. We then evaluate Eq.~\eqref{eq:BEC_TDDFPT} by Fourier interpolation of the real-space matrix elements on a non-uniform k-grid with local resolution around the Fermi surface, corresponding to a uniform $1448\times1448\times100$ grid.   
The high-resolution area in the momentum mesh must be large enough to account for the Fermi-surface shift due to the small phonon displacements of the \eg mode, which acts as a pseudo-gauge field. For a maximum phonon displacement of $0.02\,\mathrm{ \AA}$ we account for a shift of $0.03 \,\mathrm{\AA}^{-1}$ which needs to be added on top of the maximum radius of the equilibrium Fermi surface, that is $0.02 \,\mathrm{\AA}^{-1}$ \cite{dresselhaus_fermi_1964}. The denominator in~\eqref{eq:BEC_TDDFPT} is minimal for the manifold of states around the Fermi surface with a transition energy in resonance with the pump frequency of approximately $0.2\, \mathrm{eV}$. Using the Fermi velocity of $v_F \approx 6 \,\mathrm{eV\cdot \AA}$ this can be estimated to be $\sim0.015 \,\mathrm{\AA}^{-1}$ away from the Fermi surface. Therefore, the k-mesh we used maintains its maximum resolution in a cylinder around the K-H line with radius $0.1 \,\mathrm{\AA}^{-1}$ and decays cubically outside of it, matching the denominator of the BEC's real part \cite{binci_first-principles_2021}. For the coarse-grid DFT, we use a $15\times 15\times 5$ grid and a norm-conserving potential.\\
In Eq.\eqref{eq:BEC_TDDFPT}, the sum over $n,m$ runs in principle over an infinite set of bands. To perform this sum numerically, we increase the total number of bands $N$ iteratively. To preserve the $D_\mathrm{6h}$ symmetry of the crystal, $N$ cannot be increased arbitrarily but only by adding sets of symmetry-related atomic orbitals. In the Wannierization, we use the 12 $sp^2$ and 4 $p_z$ atomic orbitals as initial guesses for the first iteration with 16 bands. From there, we add ghost states at the 4 interlayer hollow sites and 2 interlayer carbon-aligned sites, adding 6 $s$-states, 18 $p$-states, and 30 $d$-states to capture graphite's quasi-free high-energy bands. Fig.~\ref{fig:lmc_displacement_fit} shows the fits for all four sets of states. Fig.~\ref{fig:lmc_coupling_convergence} shows the (dimensionless) couplings obtained from the fit using the respective number of bands. While the imaginary parts converge after $N = 40$, the real parts, particularly $c_2^{(a)}$, are not fully converged at $N = 70$. Beyond the steep numerical cost of adding more bands, the quality of disentanglement and localization in Wannier90 degrades as the number of bands increases. Because Wannierization at $N=70$ is already unreliable, we decided not to increase the number of bands further. To address this issue, we likely need a different strategy for obtaining the dynBECs.\\
As in the phonon-phonon coupling derivation the final coupling parameters used in calculations are not the values we got from DFT in units of inverse \AA$^{(2)}$, but couplings between mass-weighted normal coordinates
\begin{equation}
    g_1 = c_1 / \sqrt{M_U M_G}\quad\text{and} \quad g_2 =  c_2 / \sqrt{M_U M_G^2}
\end{equation}
in units of meV$\,$\AA\,  and (meV)$^{3/2}\,$\AA\, respectively.
In the simplified model used for the nonequilibrium theory, we do not distinguish between the polarizations. The $g_2$ coupling parameter used in those calculations is therefore set to the sum (we take the average but also multiply by two to account for the factor-$1/2$ in~\eqref{eq:Vlmc_simplified}) of $g_2^{(a)}$ and $g_2^{(b)}$. We use
\begin{subequations}
\begin{align}
\abs{\bar g_1} &= 0.013 \quad\text{and}\quad \theta_1 = 0.05\pi\\
    \abs{\bar g_2} &= 0.001\quad\text{and}\quad \theta_2 = 0.7\pi.
\end{align}
\end{subequations}
However, it should be noted that the $16$ and $22$ state calculations would imply significantly distinct $\abs{\bar g_2} = 0.003$ and $\theta_2 = 0.9\pi$. For comparison the unperturbed \eu BEC has $\abs{\bar d_\parallel} = 0.024$ and $\theta_* = 0.11\pi$. As a benchmark for the independent particle approximation, we compare the values we obtain for $\tilde d_\parallel = Z^*_\parallel/2$ using different numbers of bands in Tab.~\ref{tab:Z_parallel} with the literature value $\tilde d_\parallel = 0.27 + 0.1i$ \cite{binci_first-principles_2021}. The values we find are substantially smaller than the literature value. It does not follow that the same is true for the couplings; however, it establishes that the coupling constants we obtain are subject to significant systematic uncertainties.
\begin{table}[htbp]
    \caption{$\tilde d_\parallel$ derived from the independent particle approximation~\eqref{eq:BEC_TDDFPT} using a sum over an increasing number of states.}
    \label{tab:Z_parallel}
\renewcommand{\arraystretch}{1.5}
    \begin{tabular*}{\linewidth}{@{} l @{\extracolsep{\fill}} c c c r @{}}
        \toprule
        States & 16 & 22 & 40 & 70\\
        \midrule
        $\tilde d_\parallel$ & $0.16 + 0.07 i$ & $0.15 + 0.07 i$ & $0.20 + 0.07 i$ & $0.18 + 0.07 i$ \\
        \bottomrule
    \end{tabular*}
\end{table}
 Nonetheless, the couplings we find have the correct complex arguments to generate hardening for blue-detuned driving and softening for red-detuned driving, however their small magnitude prohibits them from causing the meV-scale resonance shifts seen in the experiment.

\section{Nonequilibrium coherent Raman scattering}\label{sec:coherent_Raman_derivation}
 In~\eqref{eq:light-matter-coupling}, we focused only on the functional dependencies and suppressed the time and space variables. More generally, the electric susceptibility acts as a convolution kernel to generate the electric dipole moment,
\begin{equation}
\begin{split}
    \vec\mu(\vec r, t) &= \int\displaylimits_{-\infty}^t d\tau \int_{r'} \chi(\vec r,\vec r';t,\tau) \cdot \vec E(\vec r',\tau) + \vec\mu_0(\vec r, t).
\end{split}
\end{equation}
IR-active modes are those that can modulate the dipole moment directly, that is, $\partial \vec\mu_0/\partial Q \neq 0$ and hence couple linearly to light. Raman active modes, on the other hand, are modulating the polarizability tensor \cite{juraschek_sum-frequency_2018},
\begin{equation}\label{eq:classical_Raman_mechanism}
    \chi_{ij}(Q) = \chi_{0,ij} + \sum_\nu \frac{\partial \chi_{ij}(Q)}{\partial Q_\nu} \Bigg\vert_{Q_\nu=0} Q_\nu + \dots\,,
\end{equation}
where $\chi_{0,ij}$ is the equilibrium value and $\nu$ sums over different phonons. In centrosymmetric crystals, these couplings are mutually exclusive, resulting in selection rules.
The derivatives $\frac{\partial \chi}{\partial Q}$ and $\frac{\partial \mu}{\partial Q}$ act as the effective light-phonon vertices, which can be obtained microscopically, for example, via DFT/TDDFPT.\\
For the coherent contribution, we can obtain a quantitative description using classical physics: A coherently oscillating Raman active phonon $Q = Q_0 \cos(\Omega_\text{ph} t)$ produces via \cref{eq:classical_Raman_mechanism} a  polarizability of the form
\begin{equation}
    \chi(t,t') = \chi_0(t-t') + \frac{\partial \chi}{\partial Q}(t-t') Q(\bar t)
\end{equation}
with the average time $\bar t = (t+t')/2$, where we approximated that the phonon is slow compared with the electronic degrees that constitute the remaining frequency dependence. Therefore, the continuous time translation invariance $\chi(t,t') = \chi(t-t')$ is broken to the discrete version $\chi(t,t') = \chi(t + T_\text{ph},t'+ T_\text{ph})$ with periodicity $T_\text{ph} = 2\pi/\Omega_\text{ph}$. The polarizability can be written as
\begin{equation}
    \chi(t,t') = \sum_{n=-1}^1 \chi_n(t-t') e^{- i n \Omega_\text{ph} \bar t}
\end{equation}
where $\chi_{\pm 1}(t-t') = \frac{Q_0}{2} \frac{\partial \chi}{\partial Q}(t-t')$. Consequently, the electric dipole moment induced by a monochromatic electric field $E(t) = E_0 \sin(\Omega_\text{in} t)$ reads
\begin{equation}\label{eq:dipole_with_side_bands}
\begin{split}
    \mu(t) &= -  \epsilon_0 E_0\Big\{ \chi_f \sin\left(\Omega_\text{in} t - \theta_f\right)\\
    &\quad + \chi_\pm  \sin\left[(\Omega_\text{in} \pm \Omega_\text{ph}) t - \theta_\pm\right]\Big\}
\end{split}
\end{equation}
with the fundamental susceptibility and phase $\chi_f = \abs{\chi_0(\Omega_\text{in})}$ and $\theta_f := \arg \chi_0(\Omega_\text{in})$, as well as the Stokes (-) and anti-Stokes (+) side-band susceptibilities and phases, $\chi_\pm := \abs{\chi_1(\Omega_\text{in}\pm \frac{\Omega_\text{ph}}{2})}$ and $\theta_\pm := \arg \chi_1\left(\Omega_\text{in}\pm \frac{\Omega_\text{ph}}{2}\right)$.\\
The oscillating dipole moment \cref{eq:dipole_with_side_bands} acts as a source term in Maxwell’s equations, which are solved in the far-field by electromagnetic radiation at the frequency of the dipole moment. If the volume or surface area in which the dipole moment is coherent significantly exceeds the wavelength of the emitted radiation, interference causes the radiation to become highly anisotropic, with the highest intensity in the forward and backward scattering directions.

\section{2PI Effective action formulation}\label{sec:2PI_derivation}
In this appendix, we derive the semiclassical theory systematically using the 2PI effective action for a scalar field theory described by an action $S = \tr \{(\partial_t \varphi)^2/2 - V[\varphi]\}$. The 2PI effective action is a functional of the mean field $\phi$ and propagator $G$ that reads \cite{berges_controlled_2002}
\begin{equation}
\begin{split}
    \Gamma_\mathrm{2PI}[\phi,G] &= S[\phi] + \frac{i}{2}\Tr \ln G\inv\\
    &\quad + \frac{i}{2} \Tr \left[ S^{(2)}[\phi] \cdot G - 1\right] + \Phi[\phi,G]\,.
\end{split}
\end{equation}
Here, the microscopic curvature is
\begin{equation}\label{eq:microscopic_curvature}
\begin{split}
    S_{ij}^{(2)}(1,2) &= \frac{\delta^2 S}{\delta \varphi_i(1) \delta \varphi_j(2)}\Big\vert_{\varphi = \phi} = -(\partial_t^2\delta_{ij} + M_{\phi,ij}^2(1)) \delta_{1,2}\\
    &\equiv G\inv_\phi \quad \text{where} \quad M_{\phi,ij}^2(1) = \frac{\partial^2 V}{\partial \varphi_i \partial \varphi_j}\Big\vert_{\varphi = \phi}\,,
\end{split} 
\end{equation}
which is the inverse of the mean/background-field propagator $G_\phi$. The Luttinger-Ward generating functional $\Phi[\phi, G]$ contains all contributions to the 2PI effective action of 2-loop order and above. For a strict tree-level (0-loop) self-energy, we must approximate $\Phi[\phi, G] \approx 0$; however, higher-order diagrams, such as the figure-eight diagram arising at the 2-loop level from a quartic interaction, that leave the self-energy time-local can be included without losing the numerical advantage of the approximation. Furthermore, coupling- and loop-order expansions are not equivalent if different vertices, e.g., cubic and quartic vertices, exist in the microscopic action, as seen in the example of the figure-eight diagram. The correlation self-energies $\Sigma^c$ are then given by
\begin{equation}
    \Sigma^c_{ij}(1,2) = 2i \frac{\delta \Phi[\phi,G]}{\delta G_{ij}(1,2)}\,.
\end{equation}
The equations of motion for $\phi$ and $G$ in the presence of a local source $J(t)$ but in the absence of a bilocal source $K(t,t')$ are given by
\begin{equation}
    \frac{\delta \Gamma_\mathrm{2PI}[\phi,G]}{\delta \phi} = J\,, \quad\text{and}\quad \frac{\delta \Gamma_\mathrm{2PI}[\phi,G]}{\delta G} = 0\,.
\end{equation}
To include the effect of external degrees of freedom, in particular acoustic phonons and electrons, we expand the action by a phenomenological embedding self-energy
\begin{equation}
    S \to S_\mathrm{open} = S - \frac{1}{2} \int_{1,2}\varphi_i(1) \Sigma_{\mathrm{em},ij}(1,2) \varphi_j(2)
\end{equation}
with the embedding rate function $\Gamma(\omega) = -2\Im \Sigma^\ret_\mathrm{em} (\omega)$ given by
\begin{equation}\label{eq:regulated_rate_function}
     \Gamma(\omega) = 2\omega\begin{pmatrix}
\gamma_\R  & 0 \\
0 & \gamma_\IR  
\end{pmatrix} \equiv 2\omega\, \Gamma\,,
\end{equation}
thus generating ohmic damping.\\
Then with $\Sigma = \Sigma^c + \Sigma_\emb$ the field EoM is
\begin{equation}
    J = \frac{\delta S[\phi]}{\delta \phi} - \Sigma_\mathrm{em} \cdot \phi + \frac{i}{2} \Tr\left[S^{(3)} G\right] + \frac{\delta \Phi[\phi,G]}{\delta \phi}
\end{equation}
and the propagator EoM is
\begin{equation}
    \left(G\inv_\phi - \Sigma\right)G = 1
\end{equation}
The KBE for the retarded and Keldysh components read
\begin{subequations}
    \begin{align}
    1 &= \left(G_\phi\inv -  \Sigma^\ret\right) \cdot G^\ret\\
    0 &= \left(G_\phi\inv - \Sigma^\ret\right) \cdot G^S - \Sigma^S \cdot G^\adv 
\end{align}\label{eq:Keldysh_KBE}
\end{subequations}
In the time domain, the embedding self-energy components take the form \cite{dolgner_hadamard_2026}
\begin{subequations}
    \begin{align}\label{eq:self-energy_A}
    \Sigma^A_\emb(t_1,t_2) &=  2\gamma \delta'(t_1-t_2)\\
    \Sigma^S_ \emb(t_1,t_2) &= \fp \left( \frac{i\gamma \pi T^2}{\sinh^2(\pi T (t_1-t_2))}\right)\label{eq:self-energy_S}
\end{align}\label{eq:time_domain_embedding_self_energy}
\end{subequations}
where f.p. denotes the Hadamard finite part. The symmetric component, generating the fluctuations, is non-local in time unless in the limit $T\to\infty$. In particular, for $T\to 0$ it decays quadratically. While methods have been developed to implement this efficiently \cite{dolgner_hadamard_2026}, taking the symmetric component at face value would spoil the time-locality that is gained uniquely in the 1-loop approximation. Instead, we apply the Markov approximation: To this end, we consider the steady state. In the frequency domain, the zero-temperature limit of the symmetric embedding self-energy takes the form
\begin{equation}
    \Sigma_\emb^S(\omega) = -i\gamma \omega \coth(\frac{\omega}{2T}) \xrightarrow{T\to 0} -i \gamma \abs{\omega}\,.
\end{equation}
In the KBEs, $\Sigma^S_\emb$ appears in a convolution with $G^\adv$, translating to a product $\Sigma^S_\emb(\omega)G^\adv(\omega)$ in the frequency domain. If we anticipate the spectral shape of $G^\adv(\omega)$ as Lorentzians centered around $\pm \omega_{\R/\IR}$ with widths $\sim\gamma_{\R/\IR}$, the embedding self-energy is effectively evaluated only at the phonon eigenfrequencies. We can approximate 
 $\Sigma^S_\emb(\omega) \approx -i\mathrm{diag}(\gamma_{\R} \omega_\R,\gamma_\IR \omega_\IR) := -iD$ if $\gamma_{\R/\IR} \ll \omega_{\R/\IR}$ such that the system effectively does not see the frequency dependence of the rate function. In the time-domain, the Markov approximation translates to a time-local diffusion kernel
$\Sigma^S_\emb(t,t') \approx -i D \delta(t-t')$, indicating a transition from quantum mechanical colored noise to classical white noise.\\
With this last approximation in place, both the retarded and Keldysh components' KBEs are time-local. We can exploit this fact by combining \eqref{eq:Keldysh_KBE} and its counterpart derived from the right-sided KBE $G\cdot (G_\phi^{-1} - \Sigma_\emb) = 1$, and rewriting them into a system of first-order ODEs for the time-diagonal, $G^S(t,t)$. Let us define 
\begin{align}
    U(t) &= iG^S(t,t)\\
    V(t) &= i\partial_{t'} G^S(t,t')\Big\vert_{t'=t}\\
    W(t) &= i\partial_{t}\partial_{t'} G^S(t,t')\Big\vert_{t'=t}
\end{align}
Then the diagonal evolves according to
\begin{subequations}
    \begin{align}
    \dot U(t) &= V^T(t) + V(t)\,,\\
    \dot V(t) &= W(t) - U(t) \cdot M_\phi^2(t) - V \cdot \Gamma\,,\\
    \begin{split}
        \dot W(t) &= - M_\phi^2(t) \cdot V(t) - V^T(t) \cdot M_\phi^2(t)\\
    &\quad - \Gamma \cdot W(t) - W(t)\cdot \Gamma + D\,.
    \end{split}
\end{align}
\label{eq:diagonal_ODE}
\end{subequations}
Here we used that $\Gamma$ and $M^2$ are symmetric matrices. For an action that is constructed from contact interactions (in time), the vertex $S^{(3)}$ is time-diagonal, that is $S_{ijk}^{(3)}(t,t',t'') = S_{ijk}^{(3)}(t)\,\delta_{t,t'}\delta_{t,t''}$. Consequently 
\begin{equation}
    \Tr(S^{(3)}\cdot G) = -i \sum_{ij}S_{ijk}^{(3)}(t) U_{ij}(t)\,.
\end{equation} 
The field EoM does not depend on the entire propagator, but only on $G^S(t,t)$:
\begin{subequations}
    \begin{align}
    J(t) &= \frac{\delta S[\phi]}{\delta \phi} - \Gamma \cdot  \dot\phi(t) + \frac{1}{2}\tr(S^{(3)}(t)\cdot U(t))
\end{align} 
\label{eq:field_ODE}
\end{subequations}

Beyond providing a systematic hierarchy of approximations that clearly contextualizes our approach, the effective action formalism offers a built-in guarantee of energy conservation because it is $\Phi$-derivable \cite{baym_conservation_1961,baym_self-consistent_1962}. In pursuit of a minimally quantum model, one might be tempted, for example, to treat the mean field entirely classically and solve for the propagator on this rigid background. However, omitting the back-action of the fluctuations on the mean field effectively makes the latter an infinite energy reservoir. Because the propagator equation of motion takes the form of a Mathieu equation, this unidirectional coupling leaves the system highly susceptible to unphysical parametric-resonance divergences.\\

The self-energies given in (\ref{eq:Gamma_R}-\ref{eq:Gamma_IRR}) can be derived using the 2PI formalism too by including only terms up to 2-loop order, that is
\begin{fmffile}{Phi-functional}
\begin{equation}
\begin{split}\label{eq:Phi-functional}
    \Phi[\phi,G] = \frac{1}{4}\quad
    \eqgraph{1ex}{0ex}{\begin{fmfgraph*}(15,15)
        \fmfleft{i}
        \fmfright{o}
        \fmf{dbl_plain,width=1,left=1}{i,o}
        \fmf{dbl_zigzag,width=1}{i,o}
        \fmf{dbl_plain,width=1,right=1}{i,o}
        \fmfdot{i,o}
    \end{fmfgraph*}} \; + \;\frac{1}{2}\quad
    \eqgraph{1ex}{0ex}{\begin{fmfgraph*}(15,15)
        \fmfleft{i}
        \fmfright{o}
        \fmftop{t}
        \fmfbottom{b}
        \fmf{dbl_plain,width=1}{i,o}
        \fmf{dbl_plain,right=0.4,tension=0.2}{i,b}
        \fmf{dbl_zigzag,right=0.4,tension=0.2}{b,o}
        \fmf{dbl_plain,left=0.4,tension=0.2}{t,o}
        \fmf{dbl_zigzag,left=0.4,tension=0.2}{i,t}
        \fmfdot{i,o}
    \end{fmfgraph*}}
    \end{split}
\end{equation}
\end{fmffile}\\

Thus, the self-energies defined in (\ref{eq:Gamma_R}-\ref{eq:Gamma_IRR}) are $\Phi$-derivable.

\end{document}